\documentclass[sn-aps,Numbered]{sn-jnl}

\usepackage{multirow}%
\usepackage{amsmath,amssymb,amsfonts}%
\usepackage{amsthm}%
\usepackage{mathrsfs}%
\usepackage[title]{appendix}%

\usepackage{textcomp}%
\usepackage{manyfoot}%
\usepackage{booktabs}%
\usepackage{algorithm}%
\usepackage{algorithmicx}%
\usepackage{algpseudocode}%
\usepackage{listings}%
\usepackage{colortbl} 
\usepackage{inputenc}

\usepackage{siunitx}
\usepackage{graphicx}
\usepackage[table,xcdraw]{xcolor}
\usepackage{adjustbox}

\begin{document}

\title[]{
Separation of Neural Drives to Muscles from Transferred Polyfunctional Nerves using Implanted Micro-electrode Arrays
}


\author*[1]{\fnm{Laura} \sur{Ferrante}}\email{l.ferrante@imperial.ac.uk}

\author[2]{\fnm{Anna} \sur{Boesendorfer}}\email{anna.boesendorfer@meduniwien.ac.at}

\author[1]{\fnm{Deren Yusuf} \sur{Barsakcioglu}}\email{deren.barsakcioglu10@imperial.ac.uk}

\author[2]{\fnm{Benedikt} \sur{Baumgartner}}\email{benedikt.baumgartner@meduniwien.ac.at}

\author[3]{\fnm{Yazan} \sur{Al-Ajam}}\email{yazan.ajam@nhs.net}

\author[3]{\fnm{Alex} \sur{Woollard}}\email{awoollard@nhs.net}

\author[3]{\fnm{Norbert Venantius} \sur{Kang}}\email{norbertkang@nhs.net}

\author*[2]{\fnm{Oskar} \sur{Aszmann}}\email{oskar.aszmann@meduniwien.ac.at}
\equalcont{Equal contribution for senior authorship}
\author*[1]{\fnm{Dario} \sur{Farina}}\email{d.farina@imperial.ac.uk}
\equalcont{Equal contribution for senior authorship}

\affil[1]{\orgdiv{Department of Bioengineering}, \orgname{Imperial College London}, \orgaddress{\city{London}, \country{United Kingdom}}}
\affil[2]{\orgdiv{Department of Plastic, Reconstructive and Aesthetic Surgery}, \orgname{Medical University of Vienna}, \orgaddress{\city{Vienna}, \country{Austria}}}
\affil[3]{\orgdiv{Department of Plastic and Reconstructive Surgery}, \orgname{Royal Free Hospital}, \orgaddress{\city{London}, \country{United Kingdom}}}

\abstract{
Following limb amputation, neural signals responsible for hand and arm functions persist in the residual peripheral nerves. Targeted muscle reinnervation (TMR) surgery allows these signals to be redirected into spare muscles, enabling the recovery of neural information through electromyography (EMG). However, a significant challenge arises in separating the distinct neural commands redirected from the transferred nerves to the muscles. Disentangling overlapping signals from EMG recordings remains complex, as they can contain mixed neural information that complicates limb function interpretation. To address this challenge, Regenerative Peripheral Nerve Interfaces (RPNIs) surgically partition the nerve into individual fascicles that reinnervate specific muscle grafts, isolating distinct neural sources for more precise control and interpretation of EMG signals. In this study, we introduce a novel biointerface that combines TMR surgery of polyvalent nerves with a high-density micro-electrode array implanted at a single site within a reinnervated muscle. Instead of surgically identifying distinct nerve fascicles, our approach separates all neural signals that are re-directed into a single muscle, using the high spatio-temporal selectivity of the micro-electrode array and mathematical source separation methods. We recorded EMG signals from four reinnervated muscles while volunteers performed phantom limb tasks. The decomposition of these signals into motor unit spike trains revealed distinct clusters of motor neurons associated with diverse functional tasks. Notably, our method enabled the extraction of multiple neural commands within a single reinnervated muscle, eliminating the need for surgical nerve division. This innovative approach not only has the potential of enhancing control over prosthetic limbs but also uncovers mechanisms of motor neuron synergies following TMR, providing valuable insights into how the central nervous system encodes movement after reinnervation.
}

\maketitle
\section{Introduction}\label{sec1}
The central nervous system controls dexterous movements of the upper limb through the coordinated activity of hundreds of thousands of motor and sensory nerve fibers. Amongst these, thousands of $\alpha$-motor neurons carry the neural signals (motor commands) from the spinal cord to the skeletal muscles. In peripheral nerves, the axons of motor neurons are organized into fascicles. Nerves can be classified based on the number of fascicles they contain \cite{stewart2003peripheral}. Monofascicular nerves have a single fascicle, oligofascicular nerves contain a few fascicles, and polyfascicular nerves have many fascicles. This classification reflects the internal organization and complexity of the nerve structure as well as the number of functions that the nerve can encode (i.e., polyvalent nerves). 
\begin{figure}[h!]
\centering
\includegraphics[width=0.9\textwidth]{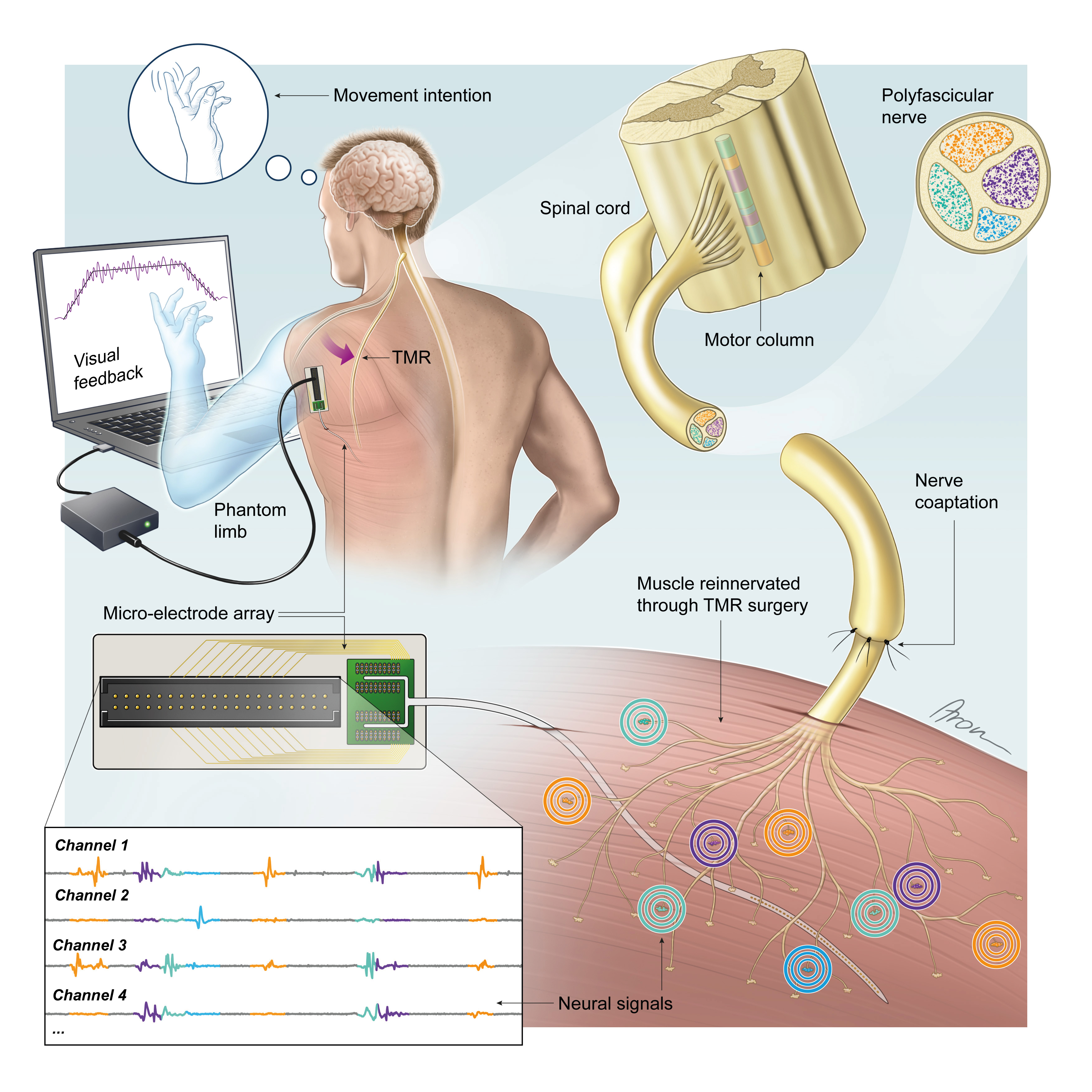}
\caption{\textbf{Biointerface based on Targeted Muscle Reinnervation (TMR) of polyfascicular nerve and the use of a single micro-electrode arrays for recording and decoding.} On the left, a glenohumeral patient has undergone TMR nerve transfer surgery: a polyfascicular nerve that previously innervated multiple upper-limb muscles (dotted purple line) was transferred into a spare target muscle (continuous light purple line). The participant is asked to perform different tasks with his phantom limb (e.g., index finger extension) while the intramuscular activity of the reinnervated muscle is recorded using a 40-channel micro-electrode array \cite{muceli2022blind}. The 40 EMG channels are distributed along 2 cm and each electrode has a diameter of \SI{140}{\micro\metre} (for comparison, muscle fibers have diameters in the range 10-\SI{100}{\micro\metre}). The participant had real-time visual feedback of the median EMG activity (purple signal) that had to be modulated to match a target trapezoidal profile displayed on the screen. This resulted in an isometric contraction of the reinnervated muscle. On the right, a schematic provides insight into the reinnervation following TMR surgery. The axons of the re-routed neurons innervate fibers of the reinnervated muscles creating a heterogeneous distribution of motor units. The activity of motor units is recorded in-vivo by the intramuscular array. From the intramuscular recordings, the individual activity of the motor units can be extracted by blind-source separation methods. In this work, we hypothesize that the heterogeneous distribution of motor units corresponds to a functional organization of motor units. Thus, clusters of motor units associated with different tasks of the phantom limb can be extracted. Note that the neural structures, muscle fibers and micro-array are not represented with an accurate scaling to improve clarity.}\label{fig:concept}
\end{figure}
Amputation of the upper extremities disrupts the upper limb motor units and thus the communication between the central nervous system and the periphery. The motor neurons previously innervating the missing limb are left without target muscles to innervate. However, neural signals for muscle control continue to travel through the nerves even after amputation. Targeted Muscle Reinnervation (TMR) involves the transfer of polyfascicular nerves, which previously innervated muscles of the missing limb, to the motor nerves of surgically denervated target muscles~\cite{kuiken2004use}. The target muscle acts as a biological amplifier for the neural signals transmitted by the re-routed nerve and allows these signals to be captured through standard human-machine interfaces using electromyography (EMG). Pioneering studies \cite{kuiken1995hyper} reported that the difference in innervation ratios of the original and targeted muscles causes the motor neurons to compete to reinnervate the available muscle fibers when the donor nerve has an axonal surplus. Although a relatively small percentage of transferred motor neurons may survive post-TMR, Bergmeister {\it{et al.}}
\cite{bergmeister2019peripheral} found no loss of muscle force generation in animal models of TMR and observed a 1.7-fold increase in the innervation ratio (i.e., hyper-reinnervation) of the targeted muscle. Subsequent histological studies demonstrated that hyper-reinnervation led to a greater number of small motor units and to a shift in the composition of the host muscle to accommodate the rerouted nerve \cite{bergmeister2021targeted}. The same authors \cite{bergmeister2017broadband} hypothesized that such hyper-reinnervation determines a heterogeneous distribution of groups of motor units that may be independently controlled, thereby potentially creating a multi-degree of freedom neural biointerface for prosthesis control.
Although a polyfascicular nerve supports extensive cognitive control and numerous functions that (in principle) can be decoded from the resulting EMG, decoding the neural signals associated with these multiple functions has proven to be a significant challenge. Classic TMR surgery transfers multiple polyfascicular nerves into distant portions of the same muscle or of different muscles to avoid interference between the recorded signals \cite{kuiken2004use,kuiken2007redirection,kuiken2009targeted,zhou2007decoding}. Surface EMG sensors are typically positioned over multiple, spatially distinct reinnervation sites, where global EMG features are analyzed to decode motor commands, generating a single control signal for each targeted reinnervation area. This contrasts with the multiple neural signals reaching the reinnervated muscle through a single polyfascicular nerve.

Regenerative Peripheral Nerve Interfaces have been proposed for enhancing signal specificity and extracting functionally separate neural signals transmitted through polyfascicular nerves. This is achieved by surgically separating the fascicles of a nerve and rerouting each one into a separate muscle graft, forming an RPNI unit \cite{vu2020regenerative}. Each RPNI unit is monitored via a bipolar intramuscular EMG. Although the donor nerve fascicles are separated without prior knowledge of their specific functional roles, a post-surgical assessment determines whether each RPNI can generate an independent EMG signal. Consequently, RPNIs offer the potential to provide multiple independent signals for controlling several degrees of freedom or tasks in a prosthetic device from a single polyfascicular nerve \cite{dumanian2009targeted,dumanian2019targeted,salminger2019outcomes,vu2020regenerative}. 

RPNIs with chronic implants have demonstrated signal stability for over a year ~\cite{vu2020regenerative,vu2023long}. However, this approach has limitations that may affect the efficacy of the nerve transfer procedure \cite{dellon2020musculus,pettersen2024regenerative}. First, the nerve is divided into fascicles without prior knowledge of the functional role of each part, which may result in RPNI units that do not provide a usable control signal. Second, there is uncertainty about whether and how much of each muscle graft will survive to create a viable RPNI unit, the surgery results in permanent damage to the donor nerve and trying to detect the EMG signals from each RPNI unit is currently impossible outside of a laboratory setting. Third, the number of fascicles - and thus the potential number of independent signals — is limited by the need to attach them to muscle grafts. Finally, these fascicles are sutured to a small segment of muscle not respecting the neuromuscular entry zone and thus run the risk of not making any functional connections.

Here, we hypothesize that neural signals carried by a polyfascicular nerve to a reinnervated muscle can be functionally separated using high-density micro-electrode arrays and a mathematical approach, avoiding the need to surgically separate nerve fascicles to create anatomically separated sites of EMG activity. With this paradigm, the vast array of functional properties of any polyvalent nerve will be displayed polytopically in a large well vascularized muscle. This would combine the benefits of both the quantity of signals with RPNI and the robust signal production with the vascularized muscle reinnervation of TMR. To evaluate this hypothesis, we recorded the high-density intramuscular EMG activity of four reinnervated muscles from three volunteers while they performed several tasks with their phantom limb, and decomposed the EMG signals into their constituent motor unit activities. This concept is illustrated in Fig.~\ref{fig:concept} for an exemplary task of index finger extension. We investigated whether the mixture of neural signals carried by the multiple fascicles of the nerve can be mathematically disentangled into a higher-dimensional neural space, from which clusters of neural activation associated with tasks of the phantom limb (e.g., pinky flexion, wrist pronation) can be extracted. Additionally, we investigated the degree of correlation in motor unit activities to formulate a conceptual framework of the synaptic input to re-formed motor units. Finally, we employed a dimensionality-reduction method to estimate the time-dependent dominant patterns of covariation in motor unit spike trains and therefore the effective dimensionality of the neural manifold underlying the muscle activity.
\begin{figure}[!]
\centering
\includegraphics[width=1\textwidth]{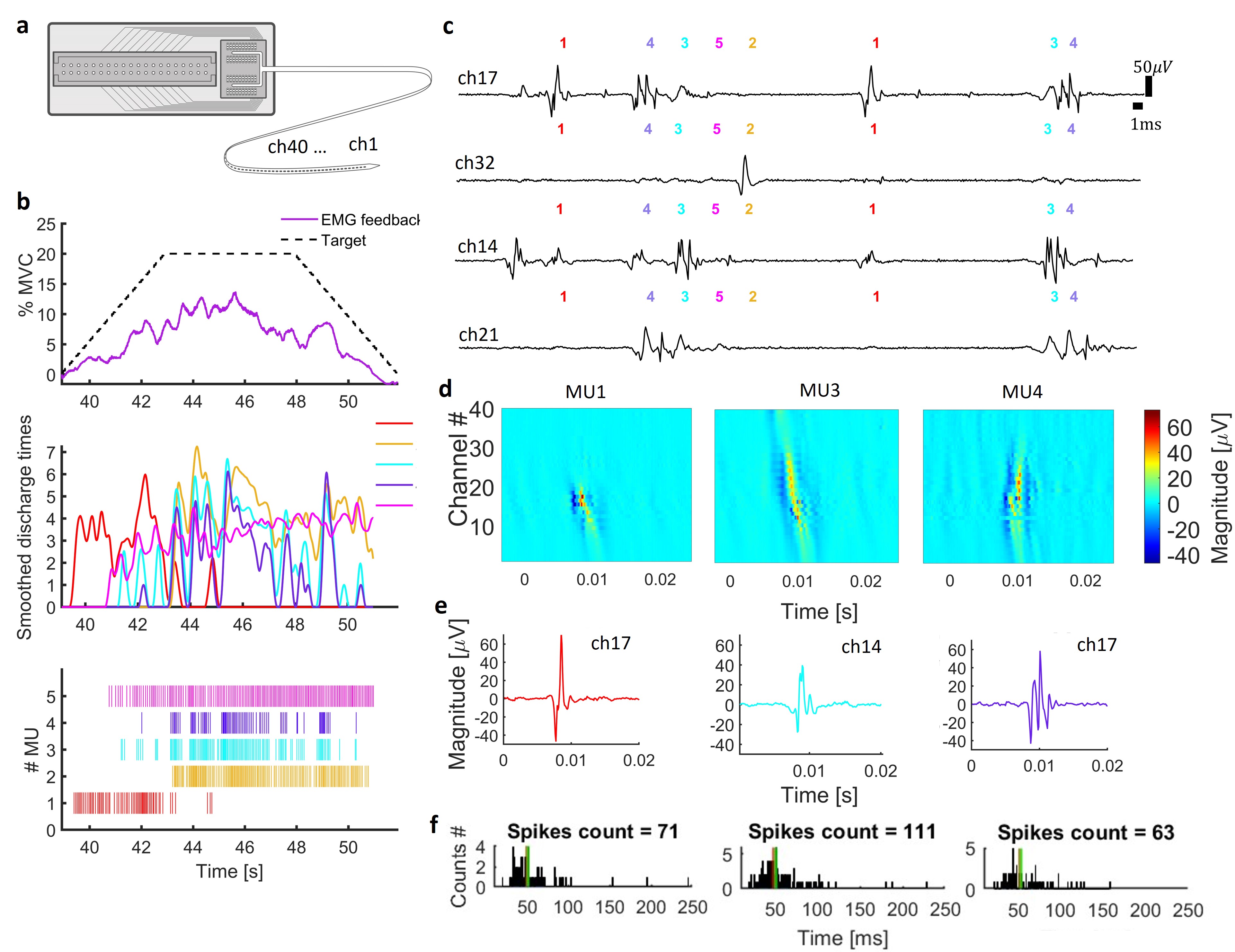}
\caption{\textbf{a}, A micro-array of EMG electrodes, containing 40 channels (ch), is implanted into each reinnervated muscle examined to record high-density intramuscular EMG activity. \textbf{b}, In an exemplary task repetition, participant P3 isometrically contracts muscle TMR4 (see Tab.~\ref{tab:patient}) while performing index finger extension with his phantom limb. A bipolar signal was derived from channels of the micro-electrode array that resulted in the maximum amplitude and was used as visual feedback of reinnervated muscle activity (purple signal). The participant modulated the muscle contraction to match the target EMG activity (dotted black line; trapezoid contraction up to a percentage of task-specific MVC). Five motor units (MUs) were reliably decomposed from the intramuscular recordings in this example. The smoothed discharges obtained by low-pass filtering the instantaneous discharge rate of each motor unit with a Hanning window of 400 ms are shown. Instantaneous discharges of active motor units are depicted with vertical lines each indicating a MU discharge at a given time instant. The smoothed and instantaneous discharges of motor units have the same color in the two plots. \textbf{c}, EMG voltages for active motor units on different channels of the micro-array. The discharge times of motor units is indicated with a color-coded number on top of each EMG signal. A channel records a MUAP with an amplitude that depends on the position of the detection site with respect to the fibers innervated by the motor unit. For this reason, the MUAP waveform differs across channels. \textbf{d}, A 2D-image of the average (across all firing instances) MUAP distribution along the 40 channels is shown for some exemplary motor units. Motor units had different morphology, as indicated by the potentials spanning fewer or almost all channels of the micro-array. \textbf{e}, Average MUAP of the motor units on the channel where the MUAP had maximum peak-to-peak amplitude. \textbf{f}, Distribution of intervals between motor unit discharges for the same motor units depicted in the panels above using 1-ms bin size.
}\label{fig:mu_properties}
\end{figure}
\section{Results}\label{sec2}
\subsection{Motor unit recordings in reinnervated muscles}\label{sec2:recordings}
Three men (aged 62, 34 and 52) who suffered from an amputation of their limb at the transhumeral (P1, P2) and glenohumeral level (P3) were included in this study. They underwent their TMR surgery at 7 to 12 years prior to the experiments, to treat neuroma and phantom limb pain (P1), and to improve prosthesis control (P2, P3). They all used a myoelectric prosthesis regularly and had good phantom sensation in their absent limbs. Detailed patient characteristics are reported in Tab.\ref{tab:patient}. A total of 4 reinnervated muscles were examined: three (TMR1, TMR2, TMR3) were reinnervated by the ulnar nerve, and TMR4 by the radial nerve.

We used a single implant for each examined reinnervated muscle to obtain a highly dense sample of intramuscular EMG activity. Each implanted micro-electrode array consisted of 40 recording sites, linearly distributed with an inter-electrode distance of \SI{500}{\micro\metre} over \SI{2}{\cm}) of overall length \cite{muceli2022blind}. Blind-source separation was used to decode the multi-channel array recordings into the activities of individual motor units. This invasive biointerface allowed us to overcome the limitations of non-invasive sensing and to obtain a high yield of decoded motor units \cite{bergmeister2021targeted}. The current study represents the first exploration of such implanted technology for a human-machine interface paradigm in TMR patients. Prior work has evaluated the feasibility of this approach in animal models of TMR \cite{muceli2018decoding}.

An example of experimental setup and protocol is shown in Fig.~\ref{fig:concept} for index finger extension performed by P3. Participants performed different tasks of their phantom limb, which resulted in contractions of their reinnervated muscle. The task-specific EMG amplitude at the Maximum Voluntary Contraction (MVC) was recorded and used to normalize subsequent EMG signals for the same task. Participants received auditory guidance from the experimenters when performing their tasks as well as real-time visual feedback of the reinnervated muscle EMG activity (as recorded by the micro-electrode array), and the targeted contraction profile they had to match. Depending on the muscle and reinnervation achieved by the TMR surgery, different phantom limb movements were included in the protocol for each participant, ranging from single (e.g., Index Finger Extension) to a combination of degrees of freedom tasks (e.g., Tripod grasp); each task was repeated 4 times by P1 and 6 times by P2 and P3. The protocol for each participant is detailed in Tab.\ref{tab:patient} and discussed in Section~\ref{sec4:sec2_protocol}. It is important to note that the EMG data for each task were normalized according to the task-specific MVC, meaning that the absolute amplitude of EMG signals at (e.g.) 20\% of MVC for a specific task, might differ from the 20\% MVC of another task. As a result, the recruitment and discharge properties of motor units varied considerably across tasks.

The micro-electrode array recorded high-quality signals in all cases (Fig.~\ref{fig:mu_properties}-a,c), with average root mean square of the baseline noise across channels below \SI{8}{\micro\volt} (5.3 $\pm$ 0.4 \SI{}{\micro\volt} TMR1, 7.8 $\pm$ 0.4 \SI{}{\micro\volt} TMR2, 5.7 $\pm$ 0.8 \SI{}{\micro\volt} TMR3, 7.8 $\pm$ 0.2 \SI{}{\micro\volt} TMR4). Fig.~\ref{fig:mu_properties}-c shows a sample of intramuscular signals from exemplary channels of the micro-electrode array. Motor unit activities can be observed from these multi-unit recordings. The EMG signals were decomposed into spike trains (Fig.~\ref{fig:mu_properties}-b) of active motor units using the blind-source-separation method presented in Muceli {\it{et al.}} \cite{muceli2022blind}, and subsequently inspected with the spike-sorting interactive software EMGLAB \cite{mcgill2005emglab} (see Sec~\ref{sec4:sec3_decom} for details).
Fig.~\ref{fig:mu_properties}-d shows examples of the distribution of the average electrical potential of the motor units across the 40 channels of the micro-electrode array, revealing motor units with different morphology and territory. In this example, MU1 spans about 10 channels, while the action potential of MU2 is distributed across tens of channels revealing a larger territory. The average values of the morphological and discharge properties of the identified motor units are reported in Tab.1 of the Supplementary Material for all the tasks performed by the reinnervated muscles.
\subsection{Motor unit activity} \label{sec2:MU_discharge_properties}
Two fundamental mechanisms underlie the coordinated control of motor units: motor units within a motor unit pool are recruited in an orderly manner \cite{denny1938fibrillation} according to the motor neuron size (i.e., the surface area of the soma and dendrites), from smaller to larger motor units \cite{henneman1957relation}; a gradual increase in force is achieved by recruiting new motor units, and concurrently by modulation of the discharge rate of the active motor units (rate coding) \cite{kukulka1981comparison,de1985control}. 
\subsubsection{Motor unit count and satellite potentials}\label{sec2:sec2:MUs_dec}
A total of 111 motor units were identified from EMG signals recorded from the four examined reinnervated muscles when considering all the tasks, with a pulse-to-noise ratio $>$\SI{30}{\decibel} \cite{holobar2014accurate}. Motor units observed in $<$ 70\% of the repetitions were not included in the motor unit count and were not used in subsequent analyses since their detection was not considered sufficiently consistent. 
The average number of identified motor units across tasks was 7.6 $\pm$ 0.8, 4.6 $\pm$ 1.1, 3.1 $\pm$ 1.4, and 3.8 $\pm$ 1.5 for the four reinnervated muscles. About a third of the motor units showed action potentials with a satellite potential, distinct from the main potential but time-locked to it \cite{lateva1999satellite}. These potentials occurred 8.8 to 32 ms from the main potential peak. A waveform was considered a satellite of an earlier occurring potential waveform if it appeared $>$ 80\% of the times concurrently with the main potential, with a constant delay. An example of satellite potential is shown in Fig.2 of the Supplementary Material. 
\subsubsection{Discharge characteristics}
\label{sec2:sec2:MU_discharge}
The distribution of the inter-spike intervals (ISI) of the motor unit discharges, during the plateau part of the contractions (shaded blue in Fig.\ref{fig:mu_properties}-b), was tested for normality. Long ISI ($>$ \SI{250}{\milli\second}) were removed as these were likely due to pauses in the tonic activity of motor unit firing. The majority of the ISI of the analysed motor units had skewness and kurtosis that deviated significantly from a normal distribution according to the D'Agostino-Pearson's test \cite{d1990suggestion}. The average percentages of normally distributed ISI considering all the units per volunteer were 27.8\% $\pm$ 14.7\%, 38.7\% $\pm$ 18.9\%, and 22.6\% $\pm$ 13.5\% for P1, P2, P3, respectively. 

We further investigated the distribution of ISIs by computing the histogram of the inter-spike intervals with 1-ms bins (Fig.\ref{fig:mu_properties}-f): ISIs had either skewed or multimodal distributions.
The median discharge rate (MFR) of the motor units was computed as the median of the inverse of the ISI of discharges during the plateau phase of the contraction.
The average task-specific MFR and the average MFR per muscle (across the different tasks) are reported in Tab.1 of the Supplementary Material. 
The average MFR across tasks was $12.39 \pm 3.26$, $13.44 \pm 3.93$, $16.12 \pm 4.23$ and $18.41 \pm 4.08$ for TMR1, TMR2, TMR3, and TMR 4, respectively. TMR1 and TMR2 had a significantly different distribution of MFR than TMR3 and TMR4 (p $<$ 0.05 in all cases); the test for TMR1 and TMR2 did not reject the null hypothesis (p = 0.3), while TMR3 and TMR4 showed a significantly different distribution of MFR (p = 0.006). 

The coefficient of variation (CoV) of the ISI, calculated as the standard deviation of the ISI divided by the median ISI, was used to measure the variability in the motor unit discharge. The average (across tasks) CoV of motor unit spike trains varied considerably across reinnervated muscles, with values ranging from $0.19 \pm 0.05\%$ (Intrinsic hand task for TMR1) to $0.54 \pm 0.10\%$ (Wrist Extension for TMR4). The average CoV across tasks was $21.72 \pm 11.30$\%, $28.50 \pm 10.65$\%, $37.35 \pm 12.90$\%, $42.84 \pm 17.58$\%, for the four reinnervated muscles, respectively. TMR1 and TMR2 distribution of CoV significantly differed from that of TMR3 and TMR4; TMR3 and TMR4 did not differ between each other (p = 0.58).
The CoV values for TMR3 and TMR4 were higher than those observed in physiologically-innervated muscles during isometric contractions, likely due to the quality of the visual feedback (EMG instead of force) and the difficulty in generating motor commands. Especially for TMR3 and TMR4, there were cases where the firing pattern of motor units exhibited irregularities (e.g., in Fig.\ref{fig:mu_properties}-b, bottom panel), such as intermittent firing throughout the isometric contraction. This might have caused an underestimation of the MFR and larger values of CoV. 
Despite large variability, the MFR were within the physiological range reported for voluntary contractions of physiologically innervated muscles (minimum 5-7 pulses per second (pps), maximum 40 pps for moderate isometric contraction ($>$ 30\% of MVC)~\cite{enoka1995morphological}.

We observed cases where later recruited motor units were the last to be de-recruited, similar to unusual motor units behaviour previously reported in elderly individuals \cite{kamen1989unusual}. An example is the Index Extension task, for TMR4, shown in Fig.\ref{fig:mu_properties}-b: MU1 is the first to be recruited, but it stops firing during the plateau phase at a higher level of MVC. MU2 is recruited during the constant phase but is de-recruited last, at a lower MVC value. These behaviours were observed consistently across repetitions of tasks.
Finally, there were cases where motor units were consistently recruited only at the beginning of the contraction (ramp-up phase) (TMR4, extension of fingers; TMR1, Intrinsic), but were not active in the other phases of the contractions.
The mean and standard deviation values of MFR and CoV per each task of the reinnervated muscles are reported in Tab.1 of the Supplementary Material.
\begin{figure}[h!]
\centering
\includegraphics[width=0.8\textwidth]{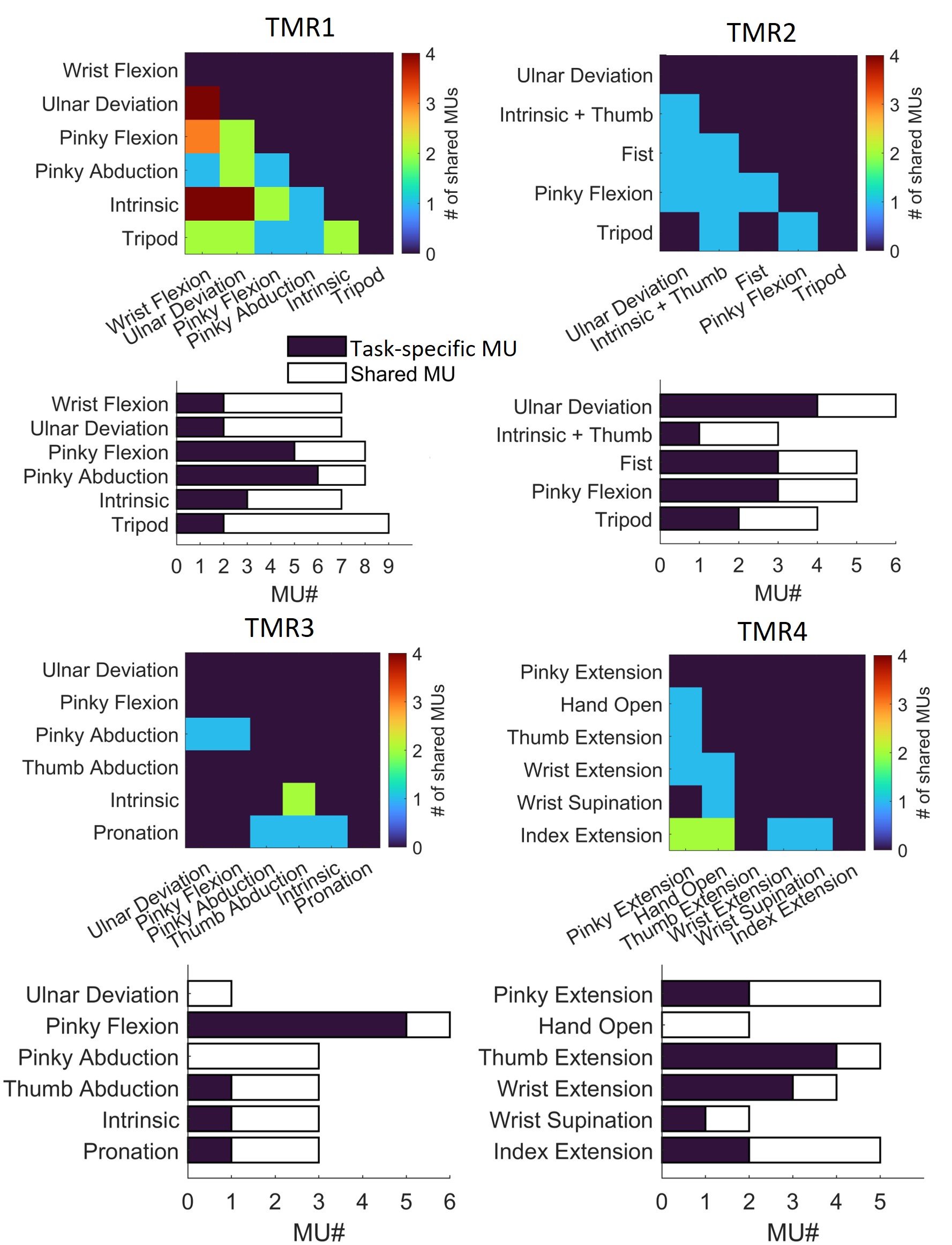}
\caption{For each reinnervated muscle, the diagrams at the top detail the relation between tasks in terms of the number of shared motor units (MUs). The bar plots report the number of shared and task-specific motor units. Only MUs that could be accurately decomposed and were active in $\geq 70\%$ of task repetitions were considered. 
}\label{fig:mu_count_shared}
\end{figure}
\subsection{Motor unit tracking}
\subsubsection{Shared and task-specific motor units}\label{sec2:sec2:MUs_shared}
For each reinnervated muscle, motor units identified during a task of the phantom limb were tracked across other tasks to determine whether they were independent (i.e., task-specific motor units) or contributed to multiple tasks (i.e., shared motor units). The average Motor Unit Action Potential (MUAP) waveform across channels obtained by spike-triggered-averaging in a \SI{20}{\milli\second} window was used to track the motor units (method detailed in Section~\ref{sec:method:mu_tracking}). The count of task-specific and shared motor units is reported in Fig.~\ref{fig:mu_count_shared}; diagrams in the same figure detail which tasks were performed with shared motor units. A total of 23 tasks were recorded considering all reinnervated muscles; 6 tasks for each of three reinnervated muscles (TMR1, TMR3 and TMR4) and 5 tasks for TMR2. Only three out of the 23 tasks lacked task-specific motor units. These were all tasks of TMR3 and TMR4 performed by the same participant P3 (Ulnar deviation, Pinky Abduction and Hand Open); only a limited number of motor units could be decomposed for these tasks (1, 3, 3 MUs, respectively).
Furthermore, none of the tasks were performed using exclusively task-specific motor units. On average, the tasks of TMR1 to TMR4 had 43.3 $\pm$ 21.2\%, 54.0 $\pm$ 13.0\%, 30.6 $\pm$ 30.6\%, 47.5 $\pm$ 28.9\% task-specific units, respectively. This analysis included only motor units recruited in at least 70\% of task repetitions.
\begin{figure}[h]
\centering
\includegraphics[width=1\textwidth]{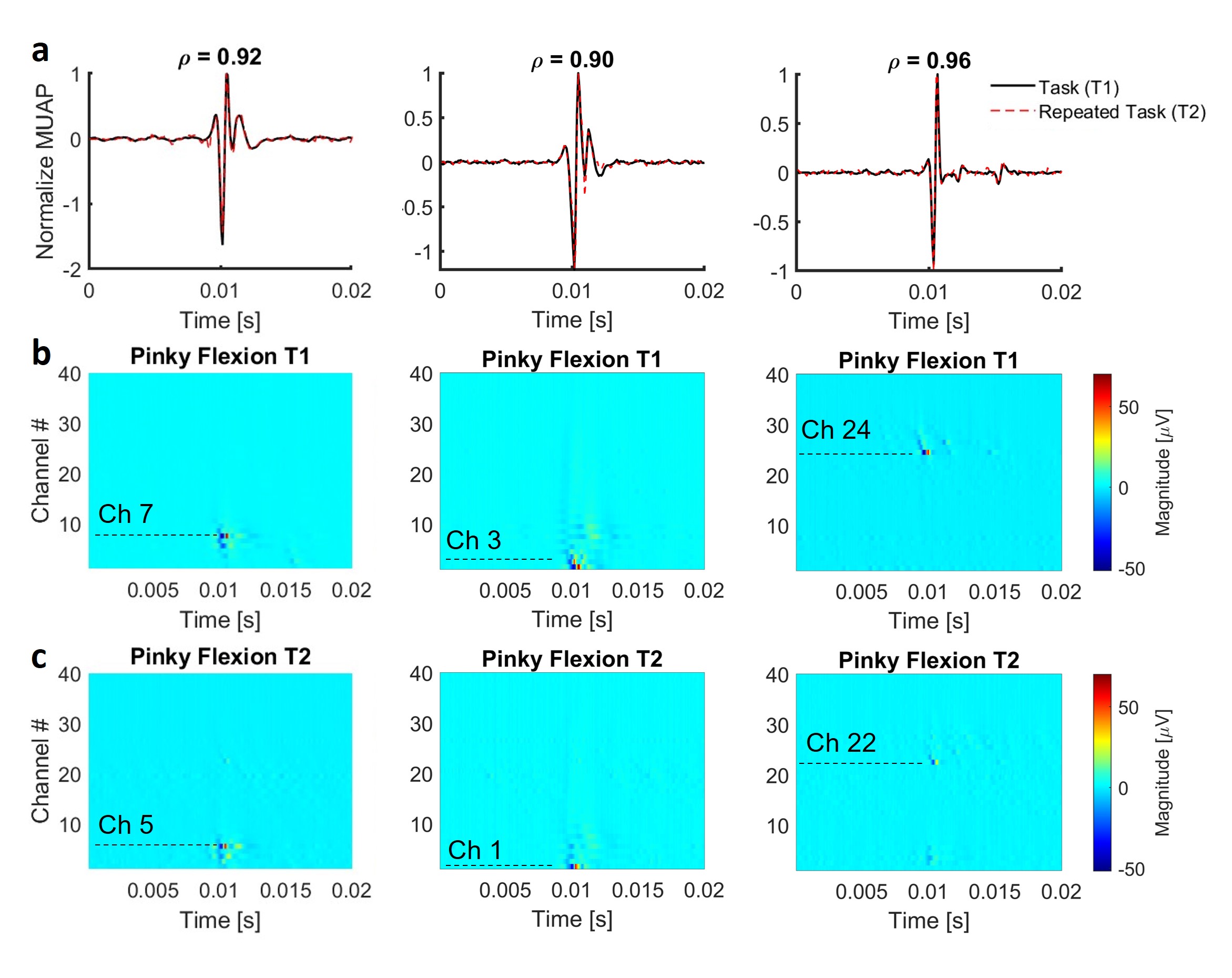}
\caption{Tracking of motor units for tasks with significant time intervals between repetitions. We show an exemplary task, Pinky Flexion, performed by participant P2 at the beginning (T1) and the end (T2) of the experimental session. \textbf{a,} EMG recordings at task repetitions T1 and T2 were decomposed separately to identify motor units. Among the five motor units recruited in T1, 3 could be tracked in T2. The normalised MUAP of the matched motor units is shown (black and red dotted line) and the goodness of the fit between the two is quantified by the coefficient of determination $\rho$. In\textbf{b,} and \textbf{c,} the distribution of motor units potential during T1 and T2 is shown, respectively. The dotted lines indicate the channel (Ch) at which the MUAPs had maximum peak-to-peak amplitude. A consistent shift of two channels can be observed: in T2 the matched motor units are shifted towards channel 1, indicating a slight micro-array displacement. The other two motor units identified during T1 could not be tracked in T2, possibly due to the electrode shift.}\label{fig:robustness}
\end{figure}
\subsubsection{Signal stability and motor unit specificity}\label{sec2:sec4_robustness}
We investigated the task-specificity and stability of the signals throughout the experimental session by asking the subjects to repeat some selected tasks (isometric contractions at the same MVC) at the beginning (T1) and end (T2) of the experimental session, which lasted a few hours. We decomposed the EMG recording of T1 and T2 separately and then checked if the same motor units could be identified during both repetitions. For example, we asked P2 to repeat Pinky Flexion and Tripod at the end of the experimental session, after repeated isometric contractions of the other tasks. We hypothesized that the same motor units could be tracked during T1 and T2 if (i) the participants could formulate the motor command for the different tasks and execute them consistently and selectively; and (ii) no significant misplacement of the electrodes had occurred. 

The micro-electrode arrays inserted in TMR2 and TMR3 were inspected to confirm no obvious displacement of the electrodes had occurred due to cable pulling or unwanted motions. As detailed in Section~\ref{sec:method:mu_tracking}, the coefficient of determination between the MUAPs $\rho \geq 0.85$ flagged a possible match between the two motor units. Additionally, the motor unit territories were estimated from the distribution of the motor unit potential across channels and compared as an additional means of tracking motor units. All pairs of motor units flagged as a match by either approach, were visually inspected by an expert examiner. These two approaches do not restrict the comparison to specific channels and can therefore track motor units even in the presence of electrode shift. Ambiguous cases, where the examined motor units had similar MUAP waveforms and were distributed in a single channel (e.g., Fig.~\ref{fig:robustness}-b) were resolved by assessing if the shift across channels was consistently detected in other tracked motor units. 
Three motor units could be tracked in T1 and T2 of Pinky Flexion. Matched motor units (in black and dashed red) are reported in Fig.~\ref{fig:robustness}-a.
In Fig.~\ref{fig:robustness}-b,c; a consistent shift of 2 channels can be observed by examining the MUAP distribution of the three motor units.
The motor unit detected in T1 had a maximum peak-to-peak amplitude at channel 7; however, this corresponded to a maximum peak-to-peak amplitude at channel 5 during T2, and the same shift was observed for the other two matched motor units. This indicated a slight displacement of the micro-electrode array along the insertion direction.

All motor units recruited during T1 of Tripod for P3, and Flexion of fingers for P2 could be identified in T2. For Pinky Flexion performed by P3, 4 out of 6 motor units were detected in both repetitions of the tasks. The two unidentified motor units had MUAPs detected only in the first few channels. 
Overall, these results suggest that the participants executed the tasks consistently and selectively as several motor units were recruited during consecutive task repetitions and, notably, also during repetitions executed after significant time intervals. 
\begin{figure}[!]
\centering
\includegraphics[width=1\textwidth]{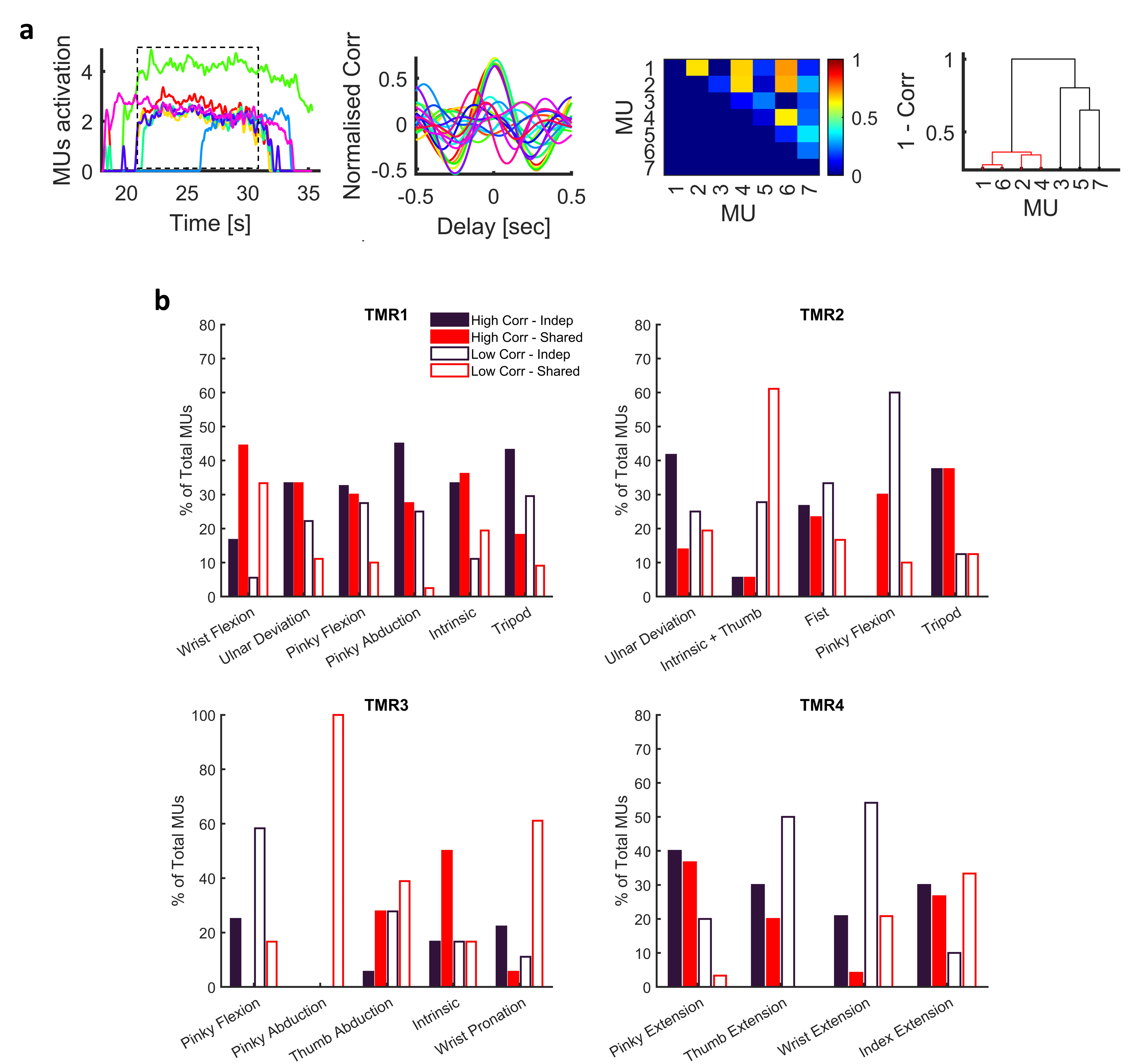}
\caption{Exemplary data of common synaptic input analysis for Intrinsic (TMR1).\textbf{a,} From left to right common oscillations in the smoothed motor units (MUs) discharges can be observed; the presence of common synaptic input to motor units is assessed by computing the cross-correlograms between smoothed and de-trended discharges of motor units (in the plateau part of the contraction); the maximum value of the cross-correlation within 100 ms of zero delay was used to quantify the strength of the common input and values are reported in the cross-correlation matrix; hierarchical clustering is applied to cluster motor units based on their inter-correlation; following the clustering analysis, motor units receiving a higher portion of common input (corr $\geq$ 0.5) are merged in a "high-correlation" cluster and "Low-correlation cluster" is formed with the MUs showing a low degree of synchronisation to other motor units (corr $<$ 0.5). \textbf{b}, For each task of the reinnervated muscles we report the proportion of task-specific and shared motor units that belonged to the high and low correlation groups.}\label{fig:common_input_model}
\end{figure}
\subsection{Estimation of neural drives from polyfascicular nerves}\label{sec2:sec3-commoninput}
\subsubsection{Correlation between motor unit activities}
Motor neurons innervating a muscle receive synaptic input from presynaptic neurons, supraspinal and afferent pathways. The net input to motor neuron comprises common components, i.e. input to multiple motor neurons, and independent ones. While common input cannot be directly measured, it causes motor units to discharge action potentials with a degree of synchrony and therefore can be inferred from the motor unit output activity by computing the cross-correlograms between the low-frequency oscillations in the motor unit spike trains \cite{de1985control,farina2016principles}. 
Based on this methodology, prior work has shown the ability of the central nervous system to selectively trigger subsets of motor units among those innervating a muscle, irrespective of anatomical constraints \cite{bremner1991correlation,bolsterlee2018three, aeles2023common,hug2023common}. 

In our data, the muscle fibers in the muscles re-innervated by polyfascicular nerves were heterogeneously activated by distinct clusters of motor neurons associated with different tasks. Within each cluster, some motor units were exclusively associated with a single task (task-specific), while others were recruited across multiple tasks (shared). To evaluate whether these motor units received common or distinct sources of synaptic input, we computed the correlation between the smoothed spike trains of pairs of recruited motor units for each task repetition. The analysis was carried out for motor units consistently active during multiple repetitions of a task, as detailed in Section~\ref{sec2:sec2:MUs_dec}. Additionally, only tasks with more than 3 motor units consistently active in $\geq$ 70\% of the repetitions were considered for this analysis. 

For each task repetition, the spike trains in the plateau phase of the isometric contraction were smoothed with a Hanning window of \SI{400}{\milli\second} to assess the common fluctuation in the low-frequency bandwidth ($<$ \SI{2.5}{\hertz}); the smoothed spike trains were subsequently high-pass filtered with a cut-off frequency of \SI{0.75}{\hertz} \cite{de2002common}. The normalised cross-correlation function was computed between pairs of filtered spike trains and the maximum cross-correlation value within \SI{100}{\milli\second} of zero-time lag was used to quantify the strength of the common input \cite{de1994common} and stored in a correlation matrix. Non-significant correlations were set to zero (Section~\ref{sec4:sec6_common}). Hierarchical clustering was then performed to group motor units based on their cross-correlation. The maximum correlation within each cluster was compared to a set threshold of 0.5 to merge clusters of motor units into either a high-correlated $G_{Corr \geq 0.5}$ or a low-correlated $G_{Corr < 0.5}$ group. These steps are illustrated in Fig.\ref{fig:common_input_model}-a for an exemplary task. 
The proportion of motor units that received common input, averaged across the tasks of each reinnervated muscle, was 65.59\%, 47.17\%, 35.69\%, and 49.58\% for TMR1 to TMR4, respectively. In Fig.\ref{fig:common_input_model}-b, we illustrate for each task the proportion of motor units that (i) were task-specific and belonged to $G_{Corr \geq 0.5}$ (blue bar); (ii) were shared and belonged to $G_{Corr \geq 0.5}$ (red bar); (iii) were task-specific and belonged to $G_{Corr < 0.5}$ (white bar with blue edge); and (iv) were shared and belonged to $G_{Corr < 0.5}$ (white bar with red edge).
The average across-tasks percentage of task-specific motor units that received common input was $34.00 \pm 10.09$\%, $25.76 \pm 18.80$\%, $18.77 \pm 15.35$\%, and $31.04 \pm 9.07$\% for TMR1 to TMR4, respectively. Instead, the proportion of shared motor units that received common input was $31.59 \pm 8.81$\%, $21.41 \pm 11.44$\%, $16.92 \pm 19.57$\%, and $18.54 \pm 12.43$\% for the four reinnervated muscles. These results indicate that all tasks involved motor units receiving common synaptic input, with the sole exception of Pinky Abduction in TMR3, where no task-specific motor units could be decomposed. Additionally, common synaptic input appeared to be distributed across both task-specific and shared motor units recruited for each task. This suggests that synergistic behavior between motor units is preserved even after reinnervation.

\begin{figure}[!]
\centering
\includegraphics[width=0.9\textwidth]{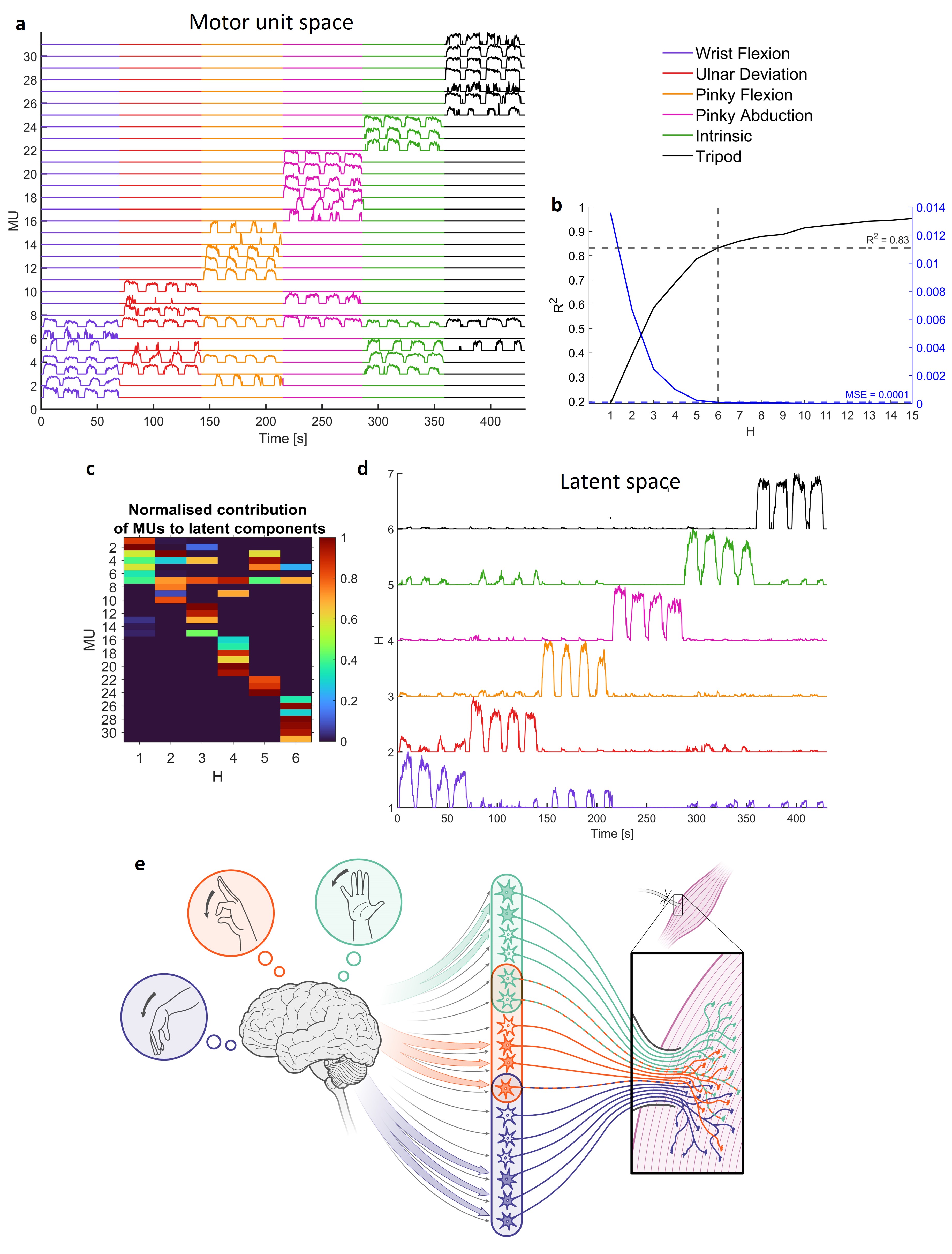}
\caption{Neural manifold analysis in muscle innervated by a poly-fascicular nerve: \textbf{a,} Smoothed motor unit (MU) spike trains from multiple tasks (colour-coded) of the phantom limb recorded from TMR1; \textbf{b,} $R^{2}$-curve obtained by applying NNMF with an increasing number of latent factors from 1 to 15 and corresponding Mean Square Error (MSE)-curve. The dashed line indicates the chosen number of latent factors. \textbf{c,} Non-negative matrix of normalised contributions of MUs to latent components output by NNMF \textbf{d,} Time-varying latent components $\textbf{H}$ embedded in the motor unit space, extracted by NNMF. Latent signals are color-coded according to the temporal distribution of activation, given the order of executed tasks indicated in (a). \textbf{e,} Conceptual model of how the central nervous system encodes movement in reinnervated muscles. Different movements of the phantom hand and their corresponding cluster of motor units are represented: wrist flexion (purple), pinky flexion (orange) and pinky abduction (green). All active motor units receive independent synaptic input (small grey arrows); some receive common synaptic input (big arrows). Within a cluster (i.e., task), motor units may be task-specific or shared between tasks (intersection between clusters). For example, there are two shared motor units for pinky flexion and abduction (orange-green) and one shared motor unit for pinky flexion and wrist flexion (orange-purple). Both task-specific or shared motor units may receive common input.}\label{fig:nmf}
\end{figure}
\subsubsection{Neural manifolds}
In the previous section, we observed that in muscles reinnervated by polyfascicular nerves, motor units recruited for a task of the phantom limb received at least a source of common synaptic input. Common inputs resulted in the observed covariation between groups of motor neuron activity. Moreover, motor neurons that receive common input may be shared across tasks. For these reasons the high-dimensional motor unit space (i.e., the space comprising the spike trains of motor units recruited across all the tasks of a reinnervated muscle) exhibits redundancy. We tested the hypothesis that the recorded neural activity of a reinnervated muscle could be modelled by a set of latent factors, with the number of factors being equal to or smaller than the number of measured tasks. We then evaluated the hypothesis that each dimension of the lower-dimensional latent space (i.e., neural manifold) captured key patterns of neural activity that corresponded to a recorded movement of the phantom limb.
For each reinnervated muscle, we examined the lower-dimensional latent space $H$ embedded in the motor unit space $X$ by applying Non-Negative Matrix Factorization (NNMF) \cite{lee2000algorithms}.

Mathematically, the problem is that of estimating two non-negative matrices $\textbf{W}$ and $\textbf{H}$ whose product approximates the original dataset $\textbf{X}$. For each reinnervated muscle, $\textbf{X}$ is obtained by sequentially concatenating the smoothed normalised spike trains of $m$ motor units active for $\geq 70\%$ of repetitions $r$ of all performed tasks $t$. An example of such an m-dimensional motor unit space $\textbf{X}$ is shown in Fig.\ref{fig:nmf}-a: the concatenated neural recordings are plotted, with temporal segments color-coded to indicate the different tasks. $\textbf{W}$ (Fig.\ref{fig:nmf}-c) is the non-negative basis matrix of dimension m (determined by total number of unique motor units measured across tasks) by $l$ (dimension of the latent space), and $\textbf{H}$ (Fig.\ref{fig:nmf}-d) contains the $l$ non-negative time-dependent latent variables. Each column of $\textbf{W}^{i}$ represents the contribution of the \textit{m} motor units to the latent signals $l$. NNMF requires the definition of the number of latent factors a priori.

\begin{figure}[H]
\centering
\includegraphics[width=1\textwidth]{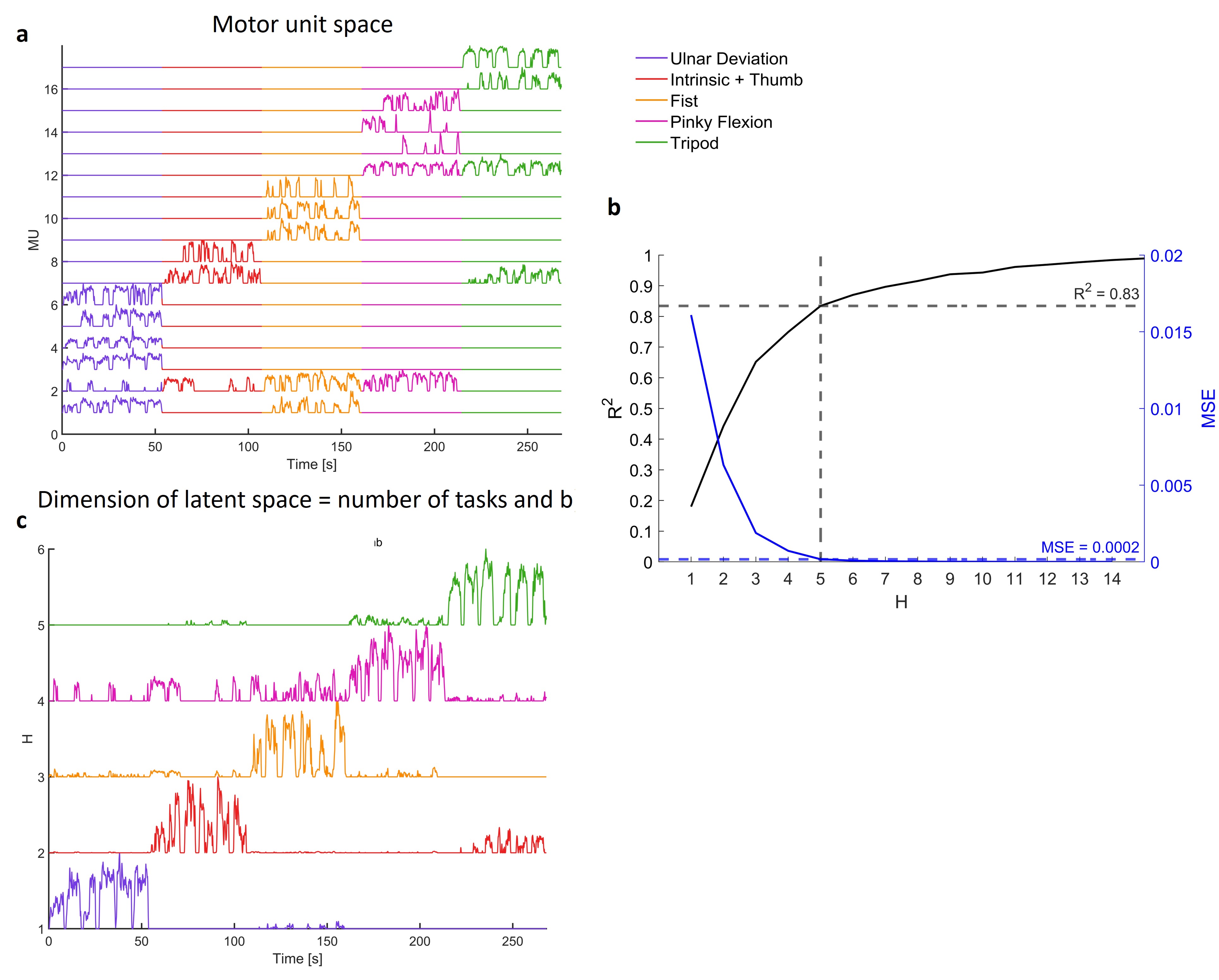}
\caption{Extended data. Exemplary data for neural manifold analysis in TMR2 muscle innervated by a polyfascicular nerve. \textbf{a,} Smoothed motor unit (MU) spike trains from multiple tasks (colour-coded) of the phantom limb recorded from TMR2 with a single 40-channel micro-electrode; the objective of the analysis is to investigate the latent space embedded in the space defined by these motor units using NNMF. \textbf{b,} $R^{2}$-curve obtained by applying NNMF with an increasing number of latent factors from 1 to 15 and corresponding Mean Square Error (MSE)-curve. The dashed line indicates the chosen number of latent factors. \textbf{c,} Time-varying latent components $\textbf{H}$ embedded in the motor unit spaced, extracted by NNMF. Latent signals are color-coded according to the temporal distribution of activation, given the order of executed tasks indicated in (a). Distinct patterns of activation corresponding to specific tasks are encoded by the latent space factors.}\label{fig:nmf3}
\end{figure}

\begin{figure}[!]
\centering
\includegraphics[width=1\textwidth]{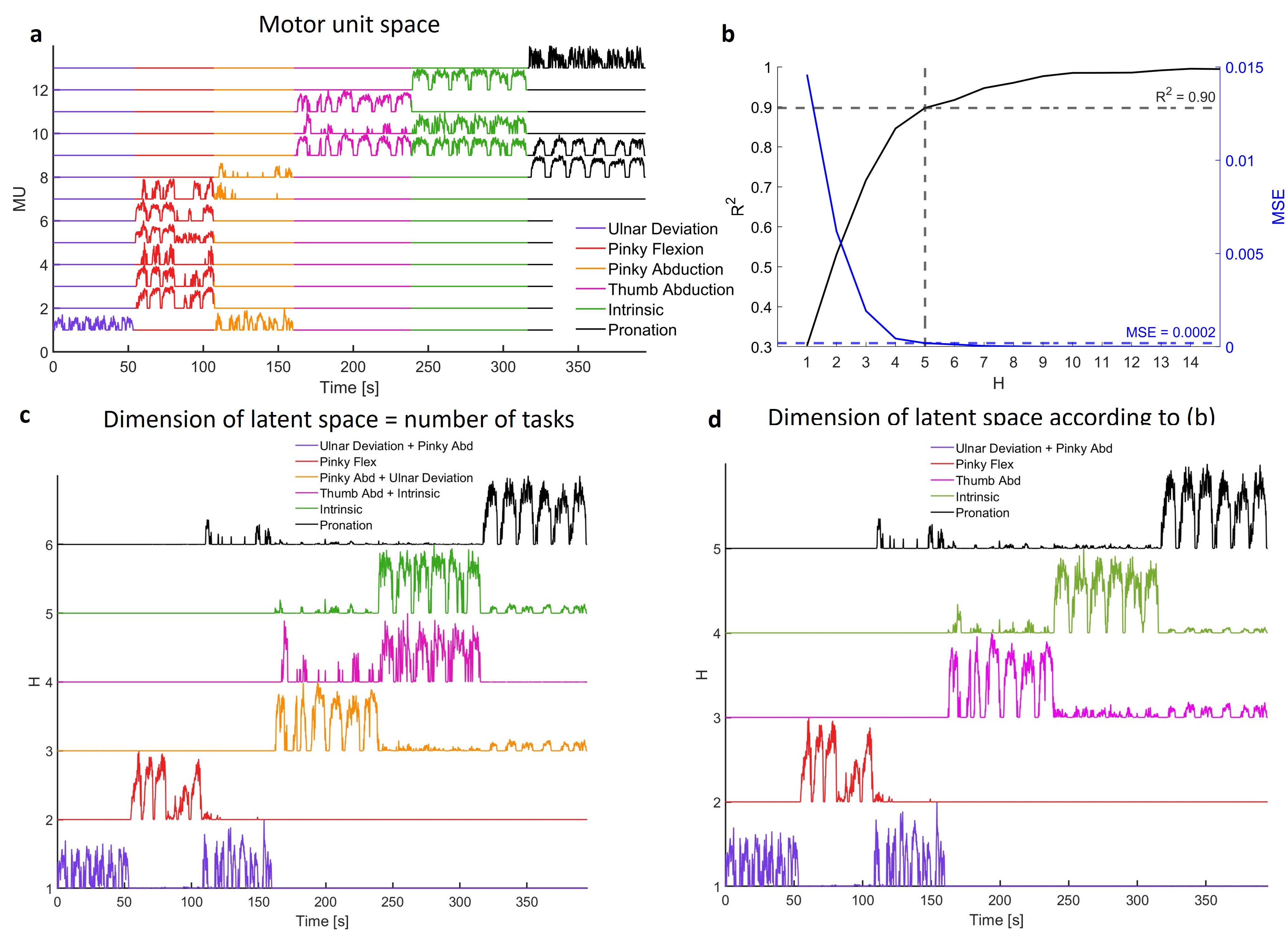}
\caption{Extended data. Exemplary data for neural manifold analysis in TMR3 muscle innervated by a polyfascicular nerve. \textbf{a,} Smoothed motor unit (MU) spike trains from multiple tasks (colour-coded) of the phantom limb recorded from TMR3 with a single 40-channel micro-electrode; the objective of the analysis is to investigate the latent space embedded in the space defined by these motor units using NNMF. \textbf{b,} $R^{2}$-curve obtained by applying NNMF with an increasing number of latent factors from 1 to 15 and corresponding Mean Square Error (MSE)-curve. The dashed line indicates the chosen number of latent factors. \textbf{c,} Time-varying latent components $\textbf{H}$ embedded in the motor unit spaced of dimension equal to the number of tasks, extracted by NNMF. Latent signals are color-coded according to the temporal distribution of activation, given the order of executed tasks indicated in (a). \textbf{d,} Time-varying latent components H embedded in the motor unit spaced of dimension estimated in (b), extracted by NNMF. Latent signals are color-coded according to the temporal distribution of activation, given the order of executed tasks indicated in (a). Distinct patterns of activation corresponding to specific tasks are encoded by the latent space factors. Task with no task-specific motor units could not be isolated.}\label{fig:nmf4}
\end{figure}

\begin{figure}[!]
\centering
\includegraphics[width=1\textwidth]{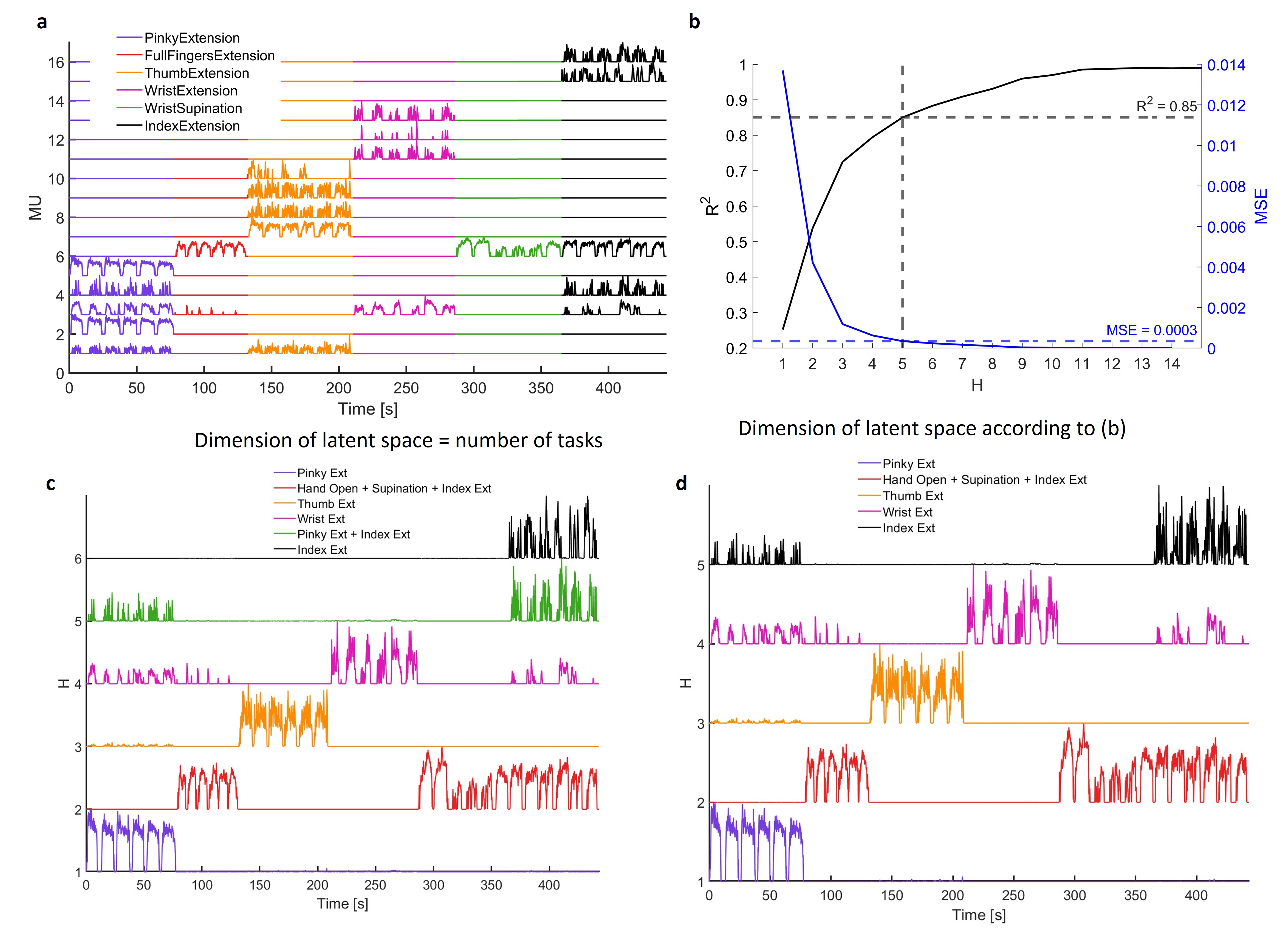}
\caption{Extended data. Exemplary data for neural manifold analysis in TMR4 muscle innervated by a polyfascicular nerve. \textbf{a,} Smoothed motor unit (MU) spike trains from multiple tasks (colour-coded) of the phantom limb recorded from TMR4 with a single 40-channel micro-electrode; the objective of the analysis is to investigate the latent space embedded in the space defined by these motor units using NNMF. \textbf{b,} $R^{2}$-curve obtained by applying NNMF with an increasing number of latent factors from 1 to 15 and corresponding Mean Square Error (MSE)-curve. The dashed line indicates the chosen number of latent factors. \textbf{c,} Time-varying latent components $\textbf{H}$ embedded in the motor unit spaced of dimension equal to the number of tasks, extracted by NNMF. Latent signals are color-coded according to the temporal distribution of activation, given the order of executed tasks indicated in (a). \textbf{d,} Time-varying latent components H embedded in the motor unit spaced of dimension estimated in (b), extracted by NNMF. Latent signals are color-coded according to the temporal distribution of activation, given the order of executed tasks indicated in (a). Distinct patterns of activation corresponding to specific tasks are encoded by the latent space factors. Task with no task-specific motor units could not be isolated.}\label{fig:nmf5}
\end{figure}

To evaluate our hypotheses we examine the latent spaces with dimensions (i) equal to the number of measured tasks and (ii) equal to the minimum number of factors beyond which an additional factor increased the coefficient of determination $R^{2}$ \cite{muceli2010identifying} between the original data X and the reconstructed data $WH$ by less than 5\% \cite{clark2010merging}. Method (ii) provided a dimensionality of the latent space that matched the number of recorded tasks in TMR1 and TMR2; a dimension of 5 was estimated for TMR3 and TMR4, which was lower than the number of tasks (6 tasks were recorded for both reinnervated muscles). Values of $R^{2}$ for an increasing number of latent factors are reported for all reinnervated muscles in Fig.\ref{fig:nmf}-b, and in the Extended Data Fig.\ref{fig:nmf3}-b, Fig.\ref{fig:nmf4}-b, and Fig.\ref{fig:nmf5}-b, respectively; a blue line indicated the obtained optimal number of factors.
All tasks in TMR1 and TMR2 had at least one task-specific motor unit, suggesting that the dimensionality of the neural manifold should correspond to the number of tasks. In contrast, since TMR3 and TMR4 had two and one task, respectively, without task-specific motor units, approach (ii) correctly predicted that the neural signals were represented by a manifold with fewer dimensions than the total number of tasks.
In panel d of Fig.\ref{fig:nmf} and in panel c of Extended Data Fig.\ref{fig:nmf3}, Fig.\ref{fig:nmf4} and Fig.\ref{fig:nmf5}, we provide the latent spaces of dimensions equal to the number of tasks; in panel d of Extended Fig.\ref{fig:nmf4} and Fig.\ref{fig:nmf5} we show the latent space estimated according to the criterion (ii), since it differs from the number of recorded tasks. Fig.\ref{fig:nmf}-d (TMR1) and Extended Data Fig.\ref{fig:nmf3}-c (TMR2) show that each dimension of the manifold had dominant activation in correspondence to a specific recorded task. Thus, the latent space dimensions effectively encoded the task-specific neural information carried by the polyfascicular nerves since at least a task-specific motor unit was active for each task. 
In TMR3, the manifold with dimension equal to the number of tasks with at least a task-specific motor unit (< total number of recorded task) allowed to separate four out of six tasks; a single latent component captured the activity of Ulnar Deviation and Pinky Abduction, which had only shared motor units. Similarly, in TMR4 the tasks with at least a task-specific motor unit could be isolated. The activity of two tasks that had only shared motor units, was captured by a single latent factor.
Overall, a total of 19 out of the 23 recorded tasks in four reinnervated muscles could be decoded from the proposed biointerface. 



\section{Discussion}\label{sec3_Discussion}
Targeted muscle reinnervation (TMR) involves rerouting of a donor nerve into a denervated muscle, allowing the muscle to receive efferent input from the donor nerve once reinnervation is complete \cite{kuiken2004use}. In this study, we investigated four reinnervated muscles in three amputee volunteers. The polyfascicular nerves, which previously carried complex motor commands for the now-absent upper limb, were surgically redirected to muscles near the stump that had lost their original function \cite{bergmeister2017broadband,bergmeister2019peripheral}. 
We demonstrated that reinnervated muscles received a diverse array of neural inputs, corresponding to the multiple distinct tasks encoded by the polyfunctional nerve, making them potentially suitable for controlling a prosthetic limb with multiple degrees of freedom. The potential for decoding multiple motor commands was demonstrated by introducing and validating a novel biointerface based on probing the reinnervated muscle connected to polyfascicular nerves, with micro-electrode array technology. With this recording approach, we captured the neural activity of individual motor units as volunteers performed various tasks with their phantom limbs. Even though we sampled intramuscular EMG activity from each reinnervated muscle at a single, small physical site (with recording channels distributed along just 2 cm), the high resolution provided by motor unit spike detection \cite{farina2017man} enabled us to differentiate neural signals corresponding to distinct tasks. These tasks ranged from fine individual finger movements to more complex grasping actions, demonstrating the biointerface's ability to decode multiple, heterogeneous neural commands from a single muscle.

We thoroughly analyzed the functional roles of motor units consistently recorded across multiple task repetitions and observed distinct clustering based on their functional contributions. In every task, some of the recruited motor units were shared across tasks. However, in 19 of the 23 tasks, at least one task-specific motor unit was uniquely recruited. These findings demonstrate that reinnervation via polyvalent nerves leads to a functional compartmentalization within the muscle. For the first time, we assessed the common synaptic input to motor neurons rerouted into a targeted muscle following TMR in human subjects. Correlation analysis of motor unit spike trains during task-specific recruitment revealed highly correlated units, indicative of shared synaptic inputs. Additionally, both task-specific and shared motor units were found to receive common synaptic inputs. These results significantly expand our understanding of the neural control of a missing limb with respect to previous studies in TMR participants that used high-density surface EMG \cite{farina2017man} and in animal models \cite{muceli2018decoding}. 

The covariation between motor units recruited for a task, and the presence of shared motor units between tasks indicated that the underlying neural activity was confined to a neural manifold of dimension lower than the number of identified motor units. We identified the neural manifold underlying the pooled data using NNMF. In TMR1 and TMR2 most of the variance within the data obtained from concatenating motor units recruited for 6 and 5 tasks, respectively, was explained by 6 and 5 latent components. Thus, each dimension of the manifold constituted a control signal for a particular task. For TMR3 and TMR4, we concatenated motor units recruited during 6 tasks for each reinnervated muscle. In both cases, the variability in the data was explained by a 5-dimensional manifold, which agrees with our preliminary analysis indicating that two tasks of TMR3 and TMR4 had only shared motor units. One could argue that motor units classified as task-specific, may not have been observed due to fluctuations in the drive to motor units. However, the reported observations were consistent across multiple repetitions of each task and for multiple motor units. Moreover, while there is no means of assessing if the participants were truly formulating the motor intent about a task, we observed consistency in motor unit recruitment across repetitions of the same task. 

The observed motor unit behavior enabled us to propose a conceptual model illustrating how the central nervous system encodes movements post-reinnervation (Fig.\ref{fig:nmf}-c). According to this model, each task is defined by a specific set of motor units, with some being uniquely recruited for a particular task (task-specific) while others are activated across multiple tasks (task-independent or shared). Both task-specific and shared motor units may receive common synaptic input, which synchronizes their activities. This pattern mirrors the control mechanisms observed in able-bodied individuals, where the central nervous system distributes common input across motor neurons, promoting modularity and reducing the complexity of control through dimensionality reduction \cite{hug2023common}. 

The findings of this study offer strong evidence of high-information transfer through TMR with polyfascicular nerves, overturning previous assumptions about the limitations of EMG in decoding information from reinnervated muscles. One such limitation is the perceived lack of isolation and specificity in control signals, stemming from the reinnervation of a single muscle by multiple nerve fascicles. Another concern is the incomplete representation of nerve functions at the innervation site, especially when the transferred nerve is at a level where fascicle organization may be disrupted, leading to an unequal distribution of encoded functions. Our results suggest that these limitations may be fully overcome when employing advanced recording and decoding techniques. 

RPNIs have been proposed as a solution to overcome the aforementioned limitations. However, RPNIs face challenges. Although previous research has shown that RPNI surgery can create functionally selective units \cite{vu2020regenerative}, there are inherent physical limitations to how many RPNIs can be generated from a single donor nerve. Furthermore, reliable techniques for extracting stable and usable EMG signals from the RPNI units for effective prosthetic control have not yet been fully established, leaving some of the key problems unresolved.

This study showcases the successful use of polyfascicular nerve transfers, combined with high-density selective recordings and advanced decoding techniques, to establish a highly efficient biointerface with the potential to control prosthetic limbs with greater functional precision and specificity. We demonstrated that diverse tasks associated with the phantom limb can be effectively decoded using a single high-density micro-electrode array. To ensure accuracy, we conducted an offline analysis, eliminating variables that could interfere with its application in real-time prosthetic control. Crucially, the principles outlined here are extendable to cases where multiple polyfascicular nerves are transferred to a single muscle, as shown in pre-clinical work by Luft {\it{et al.}} \cite{luft2021proof}. Our long-term objective is to create a hyper-reinnervated muscle, capable of replicating the full neural activity of the missing limb by integrating inputs from multiple polyfascicular nerves. This study supports prior findings in animal models \cite{muceli2018decoding}, confirming the feasibility of polyfascicular nerve transfers for sophisticated prosthetic control. This approach opens new possibilities for more advanced prosthetic solutions in the future, providing a foundation for enhanced motor control and functional specificity.
\backmatter
\section{Methods}\label{sec4}
\subsection{Subjects and nerve transfer}\label{sec4:sec1_subjects}
\begin{table}[h!]
  \centering
  \begin{adjustbox}{max width=\textwidth}
   \begin{tabular}{llll}
\toprule
{\color[HTML]{333333} \textbf{TMR volunteers}} &
  {\color[HTML]{333333} \textbf{P1}} &
  {\color[HTML]{333333} \textbf{P2}} &
  {\color[HTML]{333333} \textbf{P3}} \\ \midrule
\rowcolor[HTML]{C3C9D4} 
\multicolumn{4}{l}{\cellcolor[HTML]{C3C9D4}Patient characteristics} \\
\rowcolor[HTML]{E1E4E9} 
\textbf{Age} &
  62 y &
  34 y&
  52 y \\
\rowcolor[HTML]{FFFFFF} 
\textbf{Time since amputation} &
  44 y &
  9 y &
  19 y \\
\rowcolor[HTML]{E1E4E9} 
\textbf{Level of amputation} &
  transhumeral &
  transhumeral &
  glenohumeral \\
\rowcolor[HTML]{C3C9D4} 
\multicolumn{4}{l}{\cellcolor[HTML]{C3C9D4}Surgery details} \\
\rowcolor[HTML]{E1E4E9} 
\textbf{Time since TMR surgery} &
  7 y &
  8 y &
  12 y \\
{\color[HTML]{333333} \textbf{TMR site - innervating nerve}} &
  {\color[HTML]{000000} \begin{tabular}[c]{@{}l@{}}Pectoralis minor - Ulnar\\ Pectoralis major abdominal pars - Median\end{tabular}} &
  {\color[HTML]{333333} \begin{tabular}[c]{@{}l@{}}Biceps short head - Ulnar\\ Brachialis - Median\\ Triceps lateral head - Split deep radial branch\\ Brachioradialis - Split deep radial branch\end{tabular}} &
  {\color[HTML]{333333} \begin{tabular}[c]{@{}l@{}}Pectoralis major clavicular pars - Musculocutaneus\\ Pectoralis minor - Ulnar\\ Pectoralis major sternocostal pars - Median pars I (lat.)\\ Pectoralis major abdomial pars - Median pars II (med.)\\ Latissimus dorsi/Teres major - Radial\end{tabular}} \\
\rowcolor[HTML]{C3C9D4} 
\multicolumn{4}{l}{\cellcolor[HTML]{C3C9D4}Prosthetic information} \\
\rowcolor[HTML]{E1E4E9} 
\textbf{Prosthetic type} &
  Myoelectric &
  Myoelectric &
  Myoelectric \\
{\color[HTML]{333333} \textbf{Prosthetic use}} &
  {\color[HTML]{333333} 15h/day} &
  {\color[HTML]{333333} 2-12h/day for 2days/week} &
  {\color[HTML]{333333} 0-12h/day for 1 to 2 days/week} \\
\rowcolor[HTML]{E1E4E9} 
\textbf{No. of used EMG sensors} &
  2 &
  4 &
  6 \\
\textbf{\begin{tabular}[c]{@{}l@{}}Movements for \\ prosthetic control\end{tabular}} &
  \begin{tabular}[c]{@{}l@{}}elbow flexion, \\ elbow extension\end{tabular} &
  \begin{tabular}[c]{@{}l@{}}tripod, fingers extension, \\ elbow flexion, elbow extension\end{tabular} &
  \begin{tabular}[c]{@{}l@{}}tripod, fingers extension, elbow extension,\\ elbow flexion, pronation and supination\end{tabular} \\
\rowcolor[HTML]{C3C9D4} 
\multicolumn{4}{l}{\cellcolor[HTML]{C3C9D4} Experimental protocol} \\
\textbf{\begin{tabular}[c]{@{}l@{}}Insertion location -\\ microelectrode 1\end{tabular}} &
  Pectoralis minor - Ulnar (TMR1) &
  Biceps short head - Ulnar (TMR2) &
  Pectoralis minor  - Ulnar (TMR3) \\
\rowcolor[HTML]{E1E4E9} 
\textbf{\begin{tabular}[c]{@{}l@{}}Movements associated \\ with microelectrode 1\end{tabular}} &
  {\color[HTML]{333333} \begin{tabular}[c]{@{}l@{}}Ulnar deviation\\ Pinky flexion\\ Pinky abduction\\ Intrinsic position (MCP flexion)\\ Fist/flexion of fingers\\ Wrist flexion\end{tabular}} &
  {\color[HTML]{333333} \begin{tabular}[c]{@{}l@{}}Ulnar deviation (-)\\ Pinky flexion\\ \\ Intrinsic and thumb adduction \\ Fist/flexion of fingers\\ \\ \end{tabular}} &
  {\color[HTML]{333333} \begin{tabular}[c]{@{}l@{}}Ulnar deviation*\\ Pinky flexion\\ Pinky abduction* \\ Intrinsic position (MCP flexion) \\ Thumb adduction\\ \\ \end{tabular}} \\
\textbf{\begin{tabular}[c]{@{}l@{}}Insertion location -\\ microelectrode 2\end{tabular}} &
   &
  &
  Latissimus dorsi - Radial (TMR4) \\
\rowcolor[HTML]{E1E4E9} 
\textbf{\begin{tabular}[c]{@{}l@{}}Movements associated \\ with microelectrode 2\end{tabular}} &
  {\color[HTML]{333333} \begin{tabular}[c]{@{}l@{}} \\ \end{tabular}} &
  {\color[HTML]{333333} \begin{tabular}[c]{@{}l@{}}   \end{tabular}}&
  {\color[HTML]{333333} \begin{tabular}[c]{@{}l@{}}Supination\\ Wrist extension (-)\\ Finger extension\\ Thumb extension\\ Index finger extension (-)\\ Pinky extension\end{tabular}} \\
{\color[HTML]{333333} \textbf{Protocol}} &
  {\color[HTML]{333333} \begin{tabular}[c]{@{}l@{}}4 reps per movement, \\ Ramps: 4 s\\ Isometric: 10 sec at 10\% MVC\end{tabular}} &
  {\color[HTML]{333333} \begin{tabular}[c]{@{}l@{}}6 reps per movement\\ Ramps: 2 s\\ Isometric: 5 s at 20\% MVC\end{tabular}} &
  {\color[HTML]{333333} \begin{tabular}[c]{@{}l@{}}6 reps per movement, \\ Ramps: 2 s\\ Isometric: 5 s at 10\% MVC\\   \\ * Ramp: 4 s \\  Isometric: 5 s at 20\% MVC\end{tabular}} \\ \bottomrule
\end{tabular}%
  \end{adjustbox}
  \caption{Patient characteristics, surgery details, prosthetic information and experimental protocol. h: hours; MCP: metacarpophalangeal joint; MVC: maximal voluntary contraction; P1: patient 1; P2: patient 2; P3: patient 3; reps: repetitions; s: seconds; TMR: targeted muscle reinnervation, y: years, (-): comment of patients that the movements were difficult to imagine for the patient; * for these movements a different protocol (see *) was applied.}
  \label{tab:patient}
\end{table}
This study was approved by the Ethical Committee of Imperial College London (reference number: 19IC5641) and performed according to the Declaration of Helsinki. Three male participants (P1, P2, P3) who suffered from an amputation of their upper extremity and also underwent Targeted Muscle Reinnervation (TMR) surgery, were included in this study. Each volunteer provided written informed consent and a physical examination of their reinnervated muscles was performed prior to participation. P1 underwent his TMR surgery at the Royal Free Hospital (7 years prior); P2 (8 years prior) and P3 (12 years prior) at the Medical University of Vienna. For P1, the ulnar nerve was transferred to the pectoralis minor muscle and the median nerve was transferred to the lower (abdominal) part of the pectoralis major muscle. To make the EMG signals from the pectoralis minor easier to detect, the muscle was released from the coracoid process and was transferred into the axilla. Standard TMR procedures for a transhumeral amputation (P2) and a glenohumeral amputation (P3) were performed \cite{salminger2019outcomes,bergmeister2021targeted}. In P2 the ulnar nerve was transferred to the short head of the biceps, the median nerve to the brachialis and the split deep radial branch to the lateral head of the triceps and brachioradialis. For P3 the musculocutaneous nerve was transferred to the clavicular part of the pectoralis major, the ulnar nerve to the pectoralis minor, the median nerve to the sternocostal and abdominal part of the pectoralis major and the radial nerve to latissimus dorsi and teres major. The procedure for the nerve-to-nerve coaptation in TMR surgery where the donor nerve is sutured to the recipient nerve is described by Pettersen {\it{et al.}} \cite{pettersen2024targeted}. The stitches are positioned centrally in the donor's nerve and further sutures secure the epineurium (donor) to the fascia and epimysium (recipient). The nerve coaptation can also be performed directly at the neuromuscular entry zone. The nerve is coapted directly to the motor nerve of the muscle and its epimysium to improve stability~\cite{dumanian2009targeted}. 
The main indication for TMR surgery was the treatment of phantom limb pain for P1 and enhanced prosthesis control for P2 and P3. All three patients used a myoelectrical prosthesis in daily life with two (P1), four (P2) and six (P3) standard surface bipolar electrodes for prosthesis control. The patient characteristics and additional details on prosthesis control are summarised in Tab.~\ref{tab:patient}. 
\subsection{Electrophysiological recordings in targeted reinnervated muscles}\label{sec4:recordings}
The intramuscular electromyographic activity of each reinnervated muscle was measured using a micro-array described by Muceli {\it{et al.}} \cite{muceli2022blind}. The micro-array included 40 platinum channels (area of \SI{5257}{\micro\metre\squared}) linearly distributed with an interelectrode distance of \SI{500}{\micro\metre} over \SI{2}{\cm} of a double-sided polyimide structure \SI{20}{\micro\metre} thick. 

The insertion point for each micro-array into the corresponding reinnervated muscle was identified as the most myoelectrically active part of the muscle through clinical examination (palpation, visual muscle contraction by performing different phantom limb movement tasks related to the nerve transferred into the reinnervated muscle) and surface EMG measurements with the MyoBoy (Ottobock Healthcare Products GmbH, Duderstadt, Germany). The following 4 reinnervated muscles were examined: for P1 the pectoralis minor innervated by the ulnar nerve (TMR1); for P2 the biceps short head innervated by the ulnar nerve (TMR2); and for P3 the pectoralis minor innervated by the ulnar nerve (TMR3) and the latissimus dorsi (TMR4) innervated by the radial nerve. After disinfecting the skin, the micro-arrays were inserted acutely into the muscle at a flat angle, using a hypodermic needle of a similar size to those used in conventional concentric needle recordings. After insertion, the needle was removed while ensuring the micro-array stood in place in the muscle. The entire insertion procedure was aided by using a portable ultrasound scanner. The micro-array was removed at the end of the experimental session. A detailed insertion procedure is reported in Section 1 of the Supplementary material. 
The EMG signals were recorded in monopolar configuration using a multichannel amplifier (Quattrocento, OT Bioelettronica, Torino, Italy) with a gain of 150 and band-pass-filtered (10-\SI{4400}{\hertz}) before being sampled at \SI{10240}{\hertz} using an A/D converter to \SI{16}{\bit}. The reference and ground electrodes were placed in areas of no significant myoelectric activity depending on the level of amputation (e.g., acromion). Each micro-array provided 40 active recording channels.
\subsection{Experimental protocol}\label{sec4:sec2_protocol} 
Each participant sat in front of a computer screen and received visual feedback on the EMG activity recorded by the micro-array in reinnervated muscle. The signal used as visual feedback was the bipolar signal derived from the micro-electrode array that resulted in maximum amplitude. The volunteer was requested to perform specific movements of their phantom limb while isometrically contracting the reinnervated muscle. The list of phantom limb movements/tasks to include in the protocol was planned for each patient individually, according to their nerve transfer matrix. At the beginning of each task, the MVC was recorded and used to normalize the EMG signals of the contractions for the particular task. The volunteer was then requested to isometrically contract the reinnervated muscle and modulate the EMG activity to accurately track a series of target trapezoidal trajectories displayed on the computer screen. Each trapezoidal trajectory consisted of (i) a positive ramp phase where the muscle contraction had to be increased up to a \% of MVC; (ii) a constant contraction phase where the muscle contraction had to be maintained at a \% of MVC; and finally (iii) a negative ramp down phase where the muscle contraction had to be decreased until the muscle was fully relaxed. For example, P1 had to (i) increase the muscle contraction from 0 to 10\% of MVC in \SI{4}{\second}; (ii) maintain the contraction level at 10\% of MVC for \SI{10}{\second}; and (iii) decrease the muscle contraction in \SI{4}{\second}. The trapezoidal task was repeated 4 times per task.
\SI{20}{\second} of rest was allocated between each contractions to minimize the fatigue for all patients.

P2 performed for each task 6 isometric contraction at 10\% of MVC for five seconds and decreased back within two seconds. For P3 six repetitions per task with two (four) seconds of rise, isometric contraction at 10\% (20\%) of MVC for five seconds and decreased back within two (four) seconds. 20 seconds of rest was allocated between contractions to minimize fatigue for all patients. The list of all included tasks per patient can be found in Tab.\ref{tab:patient}. 

\subsection{Signal processing of EMG signals and quality assessment}
The recorded intramuscular EMG signals were high-pass filtered at \SI{1000}{\hertz} with a zero-lag first-order digital filter.
The quality of the signals was assessed by computing the root-mean-square of \SI{5}{\second} of data recorded at rest before starting the trials. Channels yielding a baseline noise $>$ \SI{15}{\micro\volt} were visually inspected and removed.

\subsection{Motor unit decomposition}\label{sec4:sec3_decom}
Each channel of the micro-array recorded an EMG signal given by the superimposition of the action potential propagating bidirectionally along the muscle fibers innervated by active motor neurons in the pick-up area of the electrode. Because fibers innervated by different motor neurons are intermingled, one channel might record the action potentials of multiple motor units. Moreover, the electrical activity of a motor units might be recorded by adjacent channels depending on the position of the electrode within the motor unit territory (i.e., the space defined by the fibers innervated by a motor neuron). Since each action potential is uniquely associated with a motor unit due to the high reliability of the neuromuscular junction \cite{wood2001safety}, the motor unit spike events can be extracted by explaining multiple observations of the motor unit activity using multi-channel electrodes (observations). The problem of decomposition is thus formulated as a blind source separation problem. 

The EMG signals were decomposed into their constituent motor unit spike trains using the algorithm described by Holobar and Zazula \cite{holobar2007multichannel} and validated by Muceli {\it{et al.}} \cite{muceli2022blind} using the same micro-arrays adopted in this study. 
For each task, the decomposition of signals recorded by an electrode was inferred using all the data from the repetitions.
The outcome of the automatic decomposition was validated by an expert investigator using EMGLAB \cite{mcgill2005emglab}. Specifically, each EMG signal was inspected to detect decomposition errors such as missing or incorrectly assigned discharges, paying attention to instances of long or short interspike intervals. Detection of superimposition of motor unit potentials was aided by the multi-channel recordings; each channel of the micro-array may sample a different part of the motor unit territory, providing a unique observation of the motor unit electrical activity. While the action potential morphology (i.e., amplitude, shape) differs across channels, the firing pattern is the same. This redundant information is used to resolve superimpositions of action potentials from multiple motor units. When the spike train of a motor unit was fully identified, EMGLAB subtracted the template of the MUAPs from the EMG signals. When the power of residual signal was comparable to the baseline noise level the decomposition was considered completed. Small potentials and potentials resulting in bursts of activation were not decomposed for lack of accuracy.
This procedure was repeated for each of the 40 channels of the electrode.
To ensure that the same motor unit was being identified throughout the repetitions, we calculated the average MUAP per repetition by spike-triggered-averaging, i.e., by averaging the EMG signal of each channel on the intervals of 20 ms centred around the motor unit discharges obtained from decomposition of a repetition. We then computed the coefficient of determination $\rho$ between the normalised MUAP templates and the ones of the other repetitions; additionally, motor units were the same across repetitions if their action potentials had maximum peak-to-peak amplitude on the same channel. 
Finally, we identified motor units that were satellite potentials of other units by computing the rate of agreement \cite{holobar2010experimental} between pairs of discharge patterns. The rate of agreement was defined as the ratio between the number of discharges that were present in both discharge patterns (common) and the sum of the number of common discharges and the number of discharges present in only one of the two discharge patterns. A tolerance of 10 samples ($<$ 1 ms) was used when identifying common discharges and accounting for propagation delays between the main potential and the satellite one.
If the rate of agreement between two motor units exceeded 80\% the motor unit with later discharges was considered a satellite potential of the first motor unit, removed from the motor unit list.
 
As a result of automatic and manual decomposition, the firing instances of the active motor units on each of the 40 channels were obtained. 

\subsection{Screening of motor units}
We defined the following inclusion criteria to retain identified motor units for further analysis: i) Motor units with few sparse firings were removed; ii) for a given task, we excluded those motor units that did not fire in at least 70\% of the task repetitions and had less than 30 firings; (iii) We consider only motor units whose spike trains were identified with a Pulse-to-Noise-Ratio $\geq 30$, a signal-to-interference metric \cite{holobar2014accurate}.

\subsection{Firing properties of motor units}\label{sec4:sec4_firing}
The following analysis was performed for each repetition of the different tasks. The histogram of the inter-spike intervals for each active motor unit was computed using 1-ms bins. Inter-spike intervals longer than \SI{250}{\milli\second} were considered as reflecting pauses in motor unit tonic activity and were removed from the inter-spike intervals of the motor unit.
The distribution of the inter-spike intervals was tested for normality using the D'Agostino-Pearson's test ($\alpha$ = 0.95) \cite{d1990suggestion} before characterising the average firing rate of motor units. The median firing rate (MFR) was computed as the median of the inter-spike intervals of motor units since these had a non-normal distribution. The variability of the motor unit discharges was quantified using the CoV [\%], computed as the ratio between the median and the standard deviation of ISIs. The smoothed firing rate for a motor unit was calculated by passing a Hanning window of \SI{0.4}{\second} over the impulse train corresponding to the firing times of that motor unit (instantaneous firing rate).
The quantitative values of the properties below are reported in the Supplementary Material. 
\subsection{Morphological properties of motor units}
\label{methods:sec5_morphology}
The territory of a motor unit is defined as the subarea of the cross-sectional area that encloses all the fibres belonging to a single motor unit. While a direct measurement of the motor unit territory is not possible in vivo, we estimate the territory by considering the distribution of the electrical potential across the intramuscular channels. The MUAP template, centred in a \SI{20}{\milli\second} window, is stored across the channels to build an image of the spatial distribution of the potential. We thus obtain an image where the pixel intensity corresponds to the potential amplitude. The absolute value is taken. The area occupied by the MUAP is segmented using a threshold of 15\% of the maximum absolute amplitude. The number of pixels occupied by the MUAP is normalised by the total number of pixels in the 2D image. 
Given the MUAP area, we estimate the duration [ms] of the MUAP and the MUAP cross-section as the maximum duration across the channels obtained by projecting the area on the time axis and the cross-section as the number of channels spanned by the MUAP.
The peak-to-peak unipolar amplitude [\SI{}{\micro\volt}] is computed on the channels where the MUAP has maximum value.
The MUAP size [\SI{}{\micro\volt} x \SI{}{\milli\second}] is computed as the product between the peak-to-peak unipolar amplitude and the duration of the MUAP.
For each of the computed parameters, mean and standard deviation are reported. 
\subsection{Motor unit tracking across tasks}
\label{sec:method:mu_tracking}
Given the list of motor units decomposed for a task of a reinnervated muscle, we assess whether those units were recruited for other tasks of the same reinnervated muscle (i.e., motor units shared across tasks). For each pair of motor units of two tasks we calculated the (i) coefficient of determination between the normalised MUAPs template obtained by spike-triggered averaging in a 20 ms window on the channel where the peak-to-peak unipolar amplitude of MUAPs was the largest; (ii) the distribution of the MUAP across channels was estimated as detailed above by concatenating the average MUAPs obtained for each channel to build an image of the spatial distribution of the MUAP potential (channels x time x motor unit amplitude). A 15\% threshold was applied to absolute amplitude of pixels and used to segment the area occupied by the MUAP. The MUAP cross-section (i.e., channels spanned by MUAP) was obtained by projecting the segmented area on the channel axis. Two MUAPs were flagged as belonging to the same motor units if $\rho \geq 0.85$ or if their cross-section overlapped; visual inspection assessed the match between two motor units.
\subsection{Common synaptic input to motor units of targeted reinnervated muscles}\label{sec4:sec6_common}
Motor unit synchronization was estimated with a commonly employed method \cite{nordstrom1992estimating} based on the computation of the cross-correlograms between pairs of motor units spike trains.
For each repetition of each movement, we computed the cross-correlograms between all pairs of low-pass filtered motor unit spike trains considering discharges in the constant force phase of the contraction. The spike trains were smoothed using a Hanning window of \SI{0.4}{\second} to retain the low-frequency oscillations in the signals ($<$ 2.5 Hz, effective drive \cite{farina2016principles}) and limiting the effect of the non-linear relationship between synaptic input and output signal \cite{farina2015common}. The smoothed spike trains were then high-pass filtered with a cut-off frequency of \SI{0.75}{\hertz} to remove the offset and trend \cite{de2002common}. 
The peak of the normalised cross-correlation function within \SI{100}{\milli\second} lag is considered as an index of the common drive \cite{de1994common} and used to create a correlation matrix. As a result, each task repetition had an associated cross-correlation matrix describing the synchronization between active motor units.
The statistical significance of each cross-correlogram (hence of each element in the cross-correlation matrices) was assessed as in Hug {\it{et al.}} \cite{hug2023common}. Each task had a correlation matrix and a corresponding significance matrix used to set to zero non-significant correlations.
Hierarchical clustering was performed to group motor units of a task according to their inter-correlation. A cut-off value of 0.5 was used to merge clusters of motor units into two clusters of high ($\geq$ 0.5) and low ($<$ 0.5) correlated motor units. This value was consistent across reinnervated muscles and coherent to the one obtained automatically when imposing the number of clusters equal to two.   
\subsection{Task-dependent latent neural manifolds}\label{sec:nmf}
For each reinnervated muscle (TMR$i$), the neural space embedded in the space defined by motor units decomposed for multiple tasks $\textbf{X}^{i}$ is estimated using NNMF. The matrix $\textbf{X}^{i}$, of dimension $m$ by ($r \times t$), is obtained by (i) concatenating the smoothed spike trains of motor units active for $\geq 70\%$ of repetitions $r$ of all performed tasks $t$ and (ii) by normalising the smoothed spike-trains to have unit variance. The variable $m$ indicates the total number of independent motor units across tasks, hence smoothed spike trains of motor units shared across tasks are located in the same row of $\textbf{X}^{i}$. According to NNMF, $\textbf{X}^{i}$ is mathematically described as follows:
\begin{equation}
\label{eq:model}
\textbf{X}^{i} \approx \textbf{W}^{i}\textbf{H}^{i}
\end{equation}
where $\textbf{W}^{i}$ is the non-negative basis matrix $\textbf{X}^{i}$ of dimension $m$ by $l$, and $\textbf{H}^{i}$ of dimension $l$ by ($r \times t$) is the non-negative matrix of latent time-dependent variables. Each column of $\textbf{W}^{i}$ represents the contribution of the m motor units to the latent signals $l$. Each raw of $\textbf{H}^{i}$ is a latent variable comprising the time-dependent input to motor units.
NNMF solves a non-convex optimisation problem (i.e., minimises the reconstruction error defined as the Euclidean distance between $\textbf{X}^{i}$ and $\textbf{W}^{i}\textbf{H}^{i}$), prone to converge to local minima. For this reason, given $l$, NNMF is repeated 10 times with random initialisation of $\textbf{W}^{i}$ and $\textbf{H}^{i}$ \cite{cheung2009stability} and random shuffling of motor units spike trains in $X_{i}$ and task repetitions concatenation order; the maximum number of iterations is set to 100.
Two latent spaces are explored by (i) imposing $l$ equal to the number of tasks $t$, and by (ii) setting $l$ equal to the minimum number of latent factors explaining a significant portion of the total variance within $\textbf{X}^{i}$. For (ii) various methods have been proposed: d'Avella {\it{et al.}} \cite{d2003combinations,clark2010merging} and Clark {\it{et al.}} \cite{clark2010merging} selected the minimum number of factors beyond which an additional factor increased the variance accounted for by the latent components less than 5\%. The variance accounted for was computed as $1 - \frac{SSE}{SST}$, where $SSE$ (sum of squared errors) was the unexplained variance and $SST$ (total sum of squares) was the total variance (of the data). Authors of \cite{cheung2005central} identify the number of factors for which the $R^{2}$-curve plateau by computing the mean square (MSE) curve obtained from linearly fitting the $R^{2}$-curve with an increasing number of factors and choosing the minimum number of factors that MSE lower than $10^{5}$, while this threshold was set to $5 \times 10^{-4}$ \cite{d2011superposition}.
In our study, NNMF is applied for values of $l$ ranging from 1 to 15. The coefficient of determination $R^{2}$ is computed as in Muceli {\it{et al.}} \cite{muceli2010identifying}. The MSE values are reported for completeness. The number of latent factors is chosen as the number of factors after which an additional one increases the $R^{2}$ values of < 5\%.

\subsection{Statistical analysis and reproducibility}\label{sec4:sec7_statistic}
Shapiro–Wilk tests were used to confirm homogeneity of variance and normal distribution of data respectively. 
Kruskal–Wallis one-way Anova analysis was undertaken since data violated parametric assumptions and used to analyse differences between reinnervated muscles for specific properties of motor units. 
The D’Agostino-Pearson’s test was used to test the inter-spike intervals distribution for normality. Statistical significance was assumed at P $\leq$ 0.05.
All data is reported as mean $\pm$ standard deviation. 

The following steps were implemented to ensure reproducibility: (i) the analysis reported in Section~\ref{sec2} uses only motor units that could be accurately decomposed and were detected in $\geq$ 70\% of the repetitions of the particular task; (ii) We used an automatic EMG decomposition algorithm extensively validated on various datasets and manually assessed the results using the spike-sorting software EMGLAB to identify and correct decomposition errors; (iii) tracking of motor units was done automatically and visually inspected by an expert examiner to ensure correctness.
\bmhead{Data availability}
The data supporting the findings of this article are included in the article and Supplementary material. Raw data are available from the corresponding authors upon reasonable request.
\bibliography{sn-bibliography}
\bmhead{Acknowledgements} This work was supported by the European Research Council Synergy Grant Natural BionicS (810346). The authors thank Aron Cserveny
(https://www.sciencevisual.at/) for conveying key concepts of the work through exceptional illustrations. The authors thank Agnes Sturma, Ivan Vujaklija and Silvia Muceli for the useful discussions at the beginning of the study.
\bmhead{Author contributions}
D.F. and O.A. conceived the study, L.F., D.Y.B, A.B., B.B. and O.A. performed the data acquisition, L.F. conducted the analysis, L.F. and D.F. interpreted the data, L.F., A.B., B.B. wrote or contributed to the first draft of the manuscript. All authors edited the manuscript for important scientific content and all approved the final version.
\bmhead{Competing interests}
The authors declare no competing interests.

\section*{Additional information} 
\bmhead{Ethics report} A serious adverse event was reported to the Ethics and Research Governance Coordinator. The incident was resolved after a week of hospitalisation, without any other complication for the participant. 
\bmhead{Supplementary Information} Supplementary document and figures. 
\bmhead{Correspondence and request for materials} should be addressed by L.F., D.F. and O.A.
\newpage
\section*{Extended data}











\clearpage
\section{Procedure for micro-electrode array insertion}\label{supl:thinfilms}
\begin{figure}[h]
\centering
\includegraphics[width=0.8\textwidth]{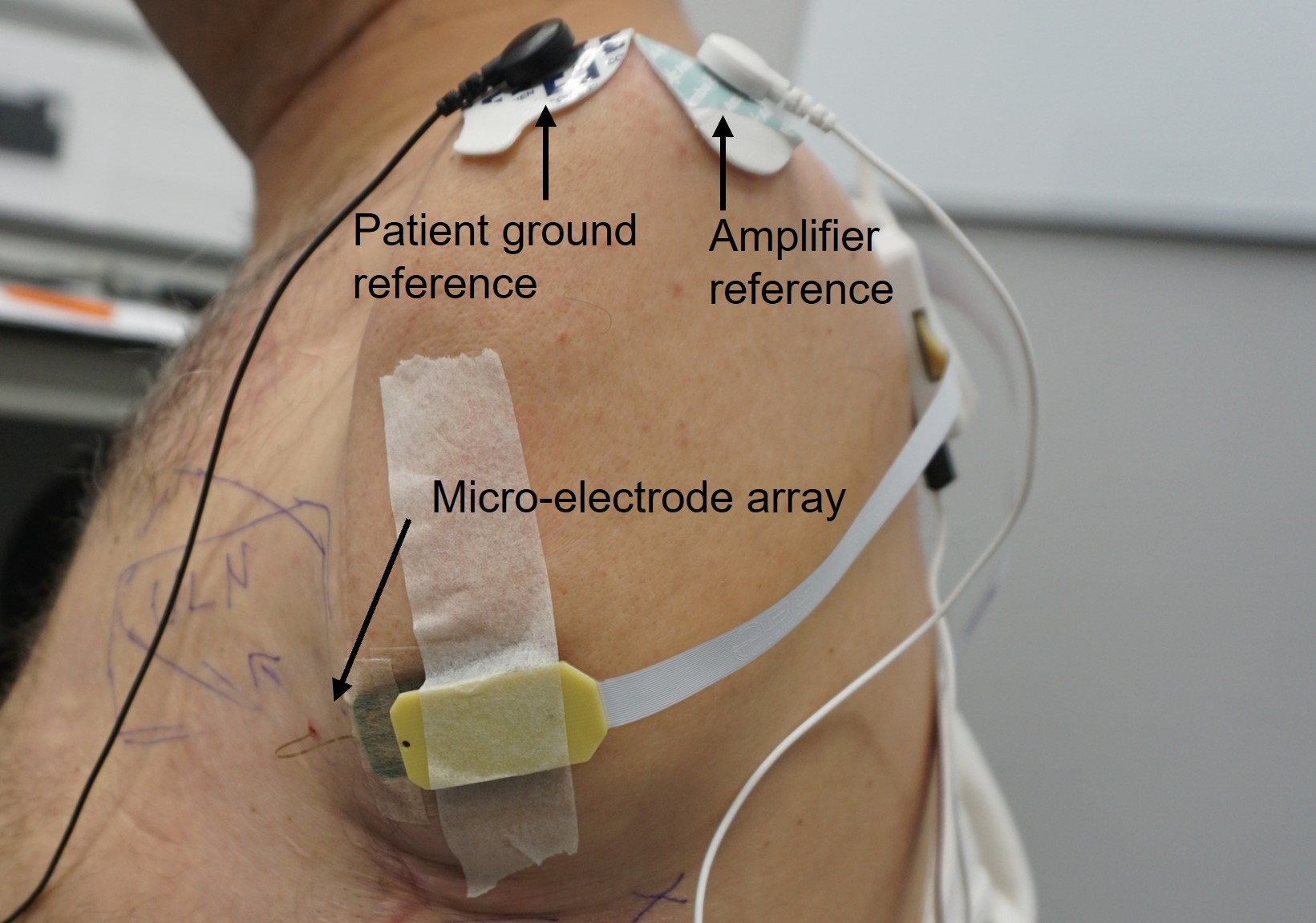}
\caption{Example of experimental setup for participant P3. The micro-electrode was inserted percutaneously in the pectoralis minor reinnervated by the ulnar nerve (TMR3). The participant ground reference and amplifier reference electrodes are placed at the acrominon.}\label{fig:thinfilms}
\end{figure}
The skin of the insertion area was disinfected before the micro-electrode array was implanted following these steps while aided by a portable ultrasound probe: (1) a hole was made with a hypodermic needle to break the skin and adipose layer barrier for smoother insertion of the micro-electrode array; (2) the package containing sterilized micro-electrode array was opened and checked to ensure none of the components were damaged; (3) the EMG needle was inserted with a flat angle (approximately 45°) through the skin into the muscle (max 2.5 to 3 cm); (4) the needle was kept inside while cutting the glue of the tiny filament; (5) the needle was removed leaving only the wire with the EMG channels inside the muscle. The electrodes were then fixed with biocompatible tapes; (6) the participant and amplifier references were placed on the acrominon for P3 or with a wet textile band on the upper arm for other participants. 
\section{Properties of motor units}
\begin{table}[!h]
  \centering
  \begin{adjustbox}{max width=\textwidth}
   \begin{tabular}{@{}lcccccc@{}}
\toprule
\multicolumn{1}{c}{\textbf{Task}} &
  \textbf{MFR {[}\SI{}{\hertz}{]}} &
  \textbf{CoV {[}\%{]}} &
  \textbf{Amplitude {[}\SI{}{\micro\volt}{]}} &
  \textbf{Duration {[}\SI{}{\milli\second}{]}} &
  \textbf{MUAP size {[}\SI{}{\micro\volt} x \SI{}{\milli\second}{]}} &
  \textbf{Normalised MUAP area} \\ \midrule
\rowcolor[HTML]{A0B2BA} 
\multicolumn{7}{|l|}{\cellcolor[HTML]{A0B2BA}\textbf{TMR1}} \\ \midrule
\multicolumn{1}{|l|}{Wrist Flexion} &
  \multicolumn{1}{c|}{11.95 $\pm$ 4.03} &
  \multicolumn{1}{c|}{0.20 $\pm$ 0.07} &
  \multicolumn{1}{c|}{52.97 $\pm$ 44.96} &
  \multicolumn{1}{c|}{1.97 $\pm$ 0.52} &
  \multicolumn{1}{c|}{96.50 $\pm$ 72.79} &
  \multicolumn{1}{c|}{0.01 $\pm$ 0.01} \\ \midrule
\multicolumn{1}{|l|}{Ulnar Deviation} &
  \multicolumn{1}{c|}{13.26 $\pm$ 3.61} &
  \multicolumn{1}{c|}{0.21 $\pm$ 0.07} &
  \multicolumn{1}{c|}{27.63 $\pm$ 24.40} &
  \multicolumn{1}{c|}{2.08 $\pm$ 0.51} &
  \multicolumn{1}{c|}{47.25 $\pm$ 26.58} &
  \multicolumn{1}{c|}{0.03 $\pm$ 0.03} \\ \midrule
\multicolumn{1}{|l|}{Pinky Flexion} &
  \multicolumn{1}{c|}{14.33 $\pm$ 4.47} &
  \multicolumn{1}{c|}{0.27 $\pm$ 0.09} &
  \multicolumn{1}{c|}{48.04 $\pm$ 42.01} &
  \multicolumn{1}{c|}{2.00 $\pm$ 0.63} &
  \multicolumn{1}{c|}{87.99 $\pm$ 90.71} &
  \multicolumn{1}{c|}{0.02 $\pm$ 0.04} \\ \midrule
\multicolumn{1}{|l|}{Pinky Abduction} &
  \multicolumn{1}{c|}{13.84 $\pm$ 3.68} &
  \multicolumn{1}{c|}{0.22 $\pm$ 0.11} &
  \multicolumn{1}{c|}{41.57 $\pm$ 23.08} &
  \multicolumn{1}{c|}{2.07 $\pm$ 0.62} &
  \multicolumn{1}{c|}{79.29 $\pm$ 44.67} &
  \multicolumn{1}{c|}{0.02 $\pm$ 0.03} \\ \midrule
\multicolumn{1}{|l|}{Intrinsic} &
  \multicolumn{1}{c|}{11.24 $\pm$ 3.69} &
  \multicolumn{1}{c|}{0.19 $\pm$ 0.05} &
  \multicolumn{1}{c|}{48.98 $\pm$ 52.52} &
  \multicolumn{1}{c|}{2.29 $\pm$ 0.86} &
  \multicolumn{1}{c|}{92.65 $\pm$ 102.06} &
  \multicolumn{1}{c|}{0.03 $\pm$ 0.04} \\ \midrule
\multicolumn{1}{|l|}{Tripod} &
  \multicolumn{1}{c|}{11.19 $\pm$ 2.38} &
  \multicolumn{1}{c|}{0.28 $\pm$ 0.08} &
  \multicolumn{1}{c|}{15.94 $\pm$ 8.99} &
  \multicolumn{1}{c|}{2.51 $\pm$ 0.76} &
  \multicolumn{1}{c|}{35.33 $\pm$ 16.48} &
  \multicolumn{1}{c|}{0.02 $\pm$ 0.01} \\ \midrule
\rowcolor[HTML]{E1E4E9} 
\multicolumn{1}{|l|}{\cellcolor[HTML]{E1E4E9}\textbf{Mean $\pm$ std}} &
  \multicolumn{1}{c|}{\cellcolor[HTML]{E1E4E9}\textbf{12.43 $\pm$ 3.46}} &
  \multicolumn{1}{c|}{\cellcolor[HTML]{E1E4E9}\textbf{0.23 $\pm$ 0.06}} &
  \multicolumn{1}{c|}{\cellcolor[HTML]{E1E4E9}\textbf{39.88 $\pm$ 39.24}} &
  \multicolumn{1}{c|}{\cellcolor[HTML]{E1E4E9}\textbf{2.18 $\pm$ 0.63}} &
  \multicolumn{1}{c|}{\cellcolor[HTML]{E1E4E9}\textbf{75.40 $\pm$ 73.06}} &
  \multicolumn{1}{c|}{\cellcolor[HTML]{E1E4E9}\textbf{0.02 $\pm$ 0.03}} \\ \midrule
\rowcolor[HTML]{A0B2BA} 
\multicolumn{7}{|l|}{\cellcolor[HTML]{A0B2BA}\textbf{TMR2}} \\ \midrule
\multicolumn{1}{|l|}{Ulnar Deviation} &
  \multicolumn{1}{l|}{12.68 $\pm$ 3.96} &
  \multicolumn{1}{l|}{0.28 $\pm$ 0.11} &
  \multicolumn{1}{l|}{145.09 $\pm$ 102.40} &
  \multicolumn{1}{l|}{1.22 $\pm$ 0.88} &
  \multicolumn{1}{l|}{119.28 $\pm$ 67.55} &
  \multicolumn{1}{l|}{0.01 $\pm$ 0.01} \\ \midrule
\multicolumn{1}{|l|}{Thumb (+ Intrinsic)} &
  \multicolumn{1}{c|}{14.16 $\pm$ 3.92} &
  \multicolumn{1}{c|}{0.33 $\pm$ 0.12} &
  \multicolumn{1}{c|}{183.92 $\pm$ 101.53} &
  \multicolumn{1}{c|}{2.03 $\pm$ 0.50} &
  \multicolumn{1}{c|}{388.38 $\pm$ 281.40} &
  \multicolumn{1}{c|}{0.01 $\pm$ 0.01} \\ \midrule
\multicolumn{1}{|l|}{Flexion of Fingers} &
  \multicolumn{1}{c|}{15.15 $\pm$ 3.03} &
  \multicolumn{1}{c|}{0.38 $\pm$ 0.12} &
  \multicolumn{1}{c|}{171.32 $\pm$ 94.21} &
  \multicolumn{1}{c|}{1.54 $\pm$ 0.94} &
  \multicolumn{1}{c|}{199.69 $\pm$ 60.89} &
  \multicolumn{1}{c|}{0.01 $\pm$ 0.01} \\ \midrule
\multicolumn{1}{|l|}{Pinky Flexion} &
  \multicolumn{1}{c|}{13.18 $\pm$ 4.16} &
  \multicolumn{1}{c|}{0.36 $\pm$ 0.12} &
  \multicolumn{1}{c|}{51.82 $\pm$ 30.76} &
  \multicolumn{1}{c|}{1.58 $\pm$ 0.45} &
  \multicolumn{1}{c|}{74.14 $\pm$ 30.50} &
  \multicolumn{1}{c|}{0.01 $\pm$ 0.01} \\ \midrule
\multicolumn{1}{|l|}{Tripod} &
  \multicolumn{1}{c|}{11.43 $\pm$ 1.91} &
  \multicolumn{1}{c|}{0.27 $\pm$ 0.07} &
  \multicolumn{1}{c|}{80.94 $\pm$ 54.33} &
  \multicolumn{1}{c|}{2.58 $\pm$ 2.21} &
  \multicolumn{1}{c|}{112.37 $\pm$ 67.10} &
  \multicolumn{1}{c|}{0.03 $\pm$ 0.04} \\ \midrule
\rowcolor[HTML]{E1E4E9} 
\multicolumn{1}{|l|}{\cellcolor[HTML]{E1E4E9}\textbf{Mean $\pm$ std}} &
  \multicolumn{1}{c|}{\cellcolor[HTML]{E1E4E9}\textbf{13.35 $\pm$ 2.89}} &
  \multicolumn{1}{c|}{\cellcolor[HTML]{E1E4E9}\textbf{0.32 $\pm$ 0.06}} &
  \multicolumn{1}{c|}{\cellcolor[HTML]{E1E4E9}\textbf{129.36 $\pm$ 91.36}} &
  \multicolumn{1}{c|}{\cellcolor[HTML]{E1E4E9}\textbf{1.72 $\pm$ 1.19}} &
  \multicolumn{1}{c|}{\cellcolor[HTML]{E1E4E9}\textbf{168.02 $\pm$ 149.62}} &
  \multicolumn{1}{c|}{\cellcolor[HTML]{E1E4E9}\textbf{0.01 $\pm$ 0.02}} \\ \midrule
\rowcolor[HTML]{A0B2BA} 
\multicolumn{7}{|l|}{\cellcolor[HTML]{A0B2BA}\textbf{TMR3}} \\ \midrule
\multicolumn{1}{|l|}{Pronation} &
  \multicolumn{1}{c|}{17.42 $\pm$ 2.88} &
  \multicolumn{1}{c|}{0.43 $\pm$ 0.20} &
  \multicolumn{1}{c|}{127.35 $\pm$ 59.02} &
  \multicolumn{1}{c|}{1.49 $\pm$ 0.67} &
  \multicolumn{1}{c|}{221.55 $\pm$ 184.60} &
  \multicolumn{1}{c|}{0.01 $\pm$ 0.01} \\ \midrule
\multicolumn{1}{|l|}{Ulnar Deviation} &
  \multicolumn{1}{c|}{10.63 $\pm$ 2.30} &
  \multicolumn{1}{c|}{0.37 $\pm$ 0.09} &
  \multicolumn{1}{c|}{18.93 $\pm$ 13.04} &
  \multicolumn{1}{c|}{3.87 $\pm$ 1.043} &
  \multicolumn{1}{c|}{76.28 $\pm$ 57.47} &
  \multicolumn{1}{c|}{0.03 $\pm$ 0.02} \\ \midrule
\multicolumn{1}{|l|}{Pinky Flexion} &
  \multicolumn{1}{c|}{14.07 $\pm$ 3.79} &
  \multicolumn{1}{c|}{0.30 $\pm$ 0.11} &
  \multicolumn{1}{c|}{73.42 $\pm$ 39.04} &
  \multicolumn{1}{c|}{2.11 $\pm$ 0.90} &
  \multicolumn{1}{c|}{144.38 $\pm$ 88.25} &
  \multicolumn{1}{c|}{0.01 $\pm$ 0.01} \\ \midrule
\multicolumn{1}{|l|}{Pinky Abduction} &
  \multicolumn{1}{c|}{13.52 $\pm$ 2.15} &
  \multicolumn{1}{c|}{0.35 $\pm$ 0.06} &
  \multicolumn{1}{c|}{16.12 $\pm$ 1.79} &
  \multicolumn{1}{c|}{4.99 $\pm$ 0.39} &
  \multicolumn{1}{c|}{80.52 $\pm$ 10.51} &
  \multicolumn{1}{c|}{0.04 $\pm$ 0.01} \\ \midrule
\multicolumn{1}{|l|}{Thumb Abduction} &
  \multicolumn{1}{c|}{20.15 $\pm$ 4.39} &
  \multicolumn{1}{c|}{0.37 $\pm$ 0.11} &
  \multicolumn{1}{c|}{88.97 $\pm$ 15.60} &
  \multicolumn{1}{c|}{1.11 $\pm$ 0.47} &
  \multicolumn{1}{c|}{94.27 $\pm$ 32.05} &
  \multicolumn{1}{c|}{0.01 $\pm$ 0.00} \\ \midrule
\multicolumn{1}{|l|}{Intrinsic} &
  \multicolumn{1}{c|}{15.85 $\pm$ 4.39} &
  \multicolumn{1}{c|}{0.39 $\pm$ 0.11} &
  \multicolumn{1}{c|}{135.28 $\pm$ 118.87} &
  \multicolumn{1}{c|}{2.43 $\pm$ 1.98} &
  \multicolumn{1}{c|}{205.28 $\pm$ 201.44} &
  \multicolumn{1}{c|}{0.02 $\pm$ 0.02} \\ \midrule
\rowcolor[HTML]{E1E4E9} 
\multicolumn{1}{|l|}{\cellcolor[HTML]{E1E4E9}\textbf{Mean $\pm$ std}} &
  \multicolumn{1}{c|}{\cellcolor[HTML]{E1E4E9}\textbf{15.33 $\pm$ 3.76}} &
  \multicolumn{1}{c|}{\cellcolor[HTML]{E1E4E9}\textbf{0.36 $\pm$ 0.09}} &
  \multicolumn{1}{c|}{\cellcolor[HTML]{E1E4E9}\textbf{83.76 $\pm$ 73.59}} &
  \multicolumn{1}{c|}{\cellcolor[HTML]{E1E4E9}\textbf{2.32 $\pm$ 1.47}} &
  \multicolumn{1}{c|}{\cellcolor[HTML]{E1E4E9}\textbf{144.57 $\pm$ 137.99}} &
  \multicolumn{1}{c|}{\cellcolor[HTML]{E1E4E9}\textbf{0.01 $\pm$ 0.01}} \\ \midrule
\rowcolor[HTML]{A0B2BA} 
\multicolumn{7}{|l|}{\cellcolor[HTML]{A0B2BA}\textbf{TMR4}} \\ \midrule
\multicolumn{1}{|l|}{Wrist Extension} &
  \multicolumn{1}{c|}{17.83 $\pm$ 1.41} &
  \multicolumn{1}{c|}{0.54 $\pm$ 0.10} &
  \multicolumn{1}{c|}{91.45 $\pm$ 20.06} &
  \multicolumn{1}{c|}{2.06 $\pm$ 0.65} &
  \multicolumn{1}{c|}{178.77 $\pm$ 31.97} &
  \multicolumn{1}{c|}{0.04 $\pm$ 0.02} \\ \midrule
\multicolumn{1}{|l|}{Supination} &
  \multicolumn{1}{c|}{19.54 $\pm$ 1.13} &
  \multicolumn{1}{c|}{0.21 $\pm$ 0.00} &
  \multicolumn{1}{c|}{32.89 $\pm$ 0.25} &
  \multicolumn{1}{c|}{2.84 $\pm$ 0.14} &
  \multicolumn{1}{c|}{93.42 $\pm$ 3.85} &
  \multicolumn{1}{c|}{0.03 $\pm$ 0.00} \\ \midrule
\multicolumn{1}{|l|}{Index Extension} &
  \multicolumn{1}{c|}{16.49 $\pm$ 2.49} &
  \multicolumn{1}{c|}{0.38 $\pm$ 0.18} &
  \multicolumn{1}{c|}{85.09 $\pm$ 46.39} &
  \multicolumn{1}{c|}{2.41 $\pm$ 0.49} &
  \multicolumn{1}{c|}{187.14 $\pm$ 68.44} &
  \multicolumn{1}{c|}{0.03 $\pm$ 0.01} \\ \midrule
\multicolumn{1}{|l|}{Pinky Extension} &
  \multicolumn{1}{c|}{21.16 $\pm$ 4.35} &
  \multicolumn{1}{c|}{0.47 $\pm$ 0.27} &
  \multicolumn{1}{c|}{66.64 $\pm$ 47.83} &
  \multicolumn{1}{c|}{2.39 $\pm$ 0.87} &
  \multicolumn{1}{c|}{146.34 $\pm$ 83.24} &
  \multicolumn{1}{c|}{0.03 $\pm$ 0.01} \\ \midrule
\multicolumn{1}{|l|}{Thumb Extension} &
  \multicolumn{1}{c|}{20.39 $\pm$ 4.14} &
  \multicolumn{1}{c|}{0.47 $\pm$ 0.23} &
  \multicolumn{1}{c|}{59.03 $\pm$ 50.21} &
  \multicolumn{1}{c|}{2.92 $\pm$ 0.81} &
  \multicolumn{1}{c|}{147.93 $\pm$ 95.72} &
  \multicolumn{1}{c|}{0.04 $\pm$ 0.02} \\ \midrule
\multicolumn{1}{|l|}{Extension of Fingers} &
  \multicolumn{1}{c|}{14.65 $\pm$ 1.38} &
  \multicolumn{1}{c|}{0.21 $\pm$ 0.07} &
  \multicolumn{1}{c|}{35.91 $\pm$ 2.31} &
  \multicolumn{1}{c|}{3.07 $\pm$ 0.25} &
  \multicolumn{1}{c|}{110.24 $\pm$ 9.13} &
  \multicolumn{1}{c|}{0.03 $\pm$ 0.01} \\ \midrule
\rowcolor[HTML]{E1E4E9} 
\multicolumn{1}{|l|}{\cellcolor[HTML]{E1E4E9}\textbf{Mean $\pm$ std}} &
  \multicolumn{1}{c|}{\cellcolor[HTML]{E1E4E9}\textbf{19.13 $\pm$ 3.44}} &
  \multicolumn{1}{c|}{\cellcolor[HTML]{E1E4E9}\textbf{0.44 $\pm$ 0.17}} &
  \multicolumn{1}{c|}{\cellcolor[HTML]{E1E4E9}\textbf{71.79 $\pm$ 45.92}} &
  \multicolumn{1}{c|}{\cellcolor[HTML]{E1E4E9}\textbf{2.54 $\pm$ 0.66}} &
  \multicolumn{1}{c|}{\cellcolor[HTML]{E1E4E9}\textbf{160.84 $\pm$ 77.65}} &
  \multicolumn{1}{c|}{\cellcolor[HTML]{E1E4E9}\textbf{0.03 $\pm$ 0.01}} \\ \bottomrule
\end{tabular}%

\end{adjustbox}
  \caption{Average values motor unit properties across tasks repetitions. Median firing rate [\SI{}{\hertz}], Peak-to-peak unipolar amplitude [\SI{}{\micro\volt}] computed on the channel where the MUAP had maximum amplitude, duration of MUAP [\SI{}{\milli\second}] computed on the channel where the MUAP had maximum duration, normalized area occupied by the MU potential across the 40 channels considering a \SI{20}{\milli\second} time window centred at the MUAP main peak. A detailed description of such properties is provided in the Methodology section of the main paper.}
  \label{tab:mu_stats}
\end{table}
\newpage
\section{Satellite potentials}
\begin{figure}[h]
\centering
\includegraphics[width=1\textwidth]{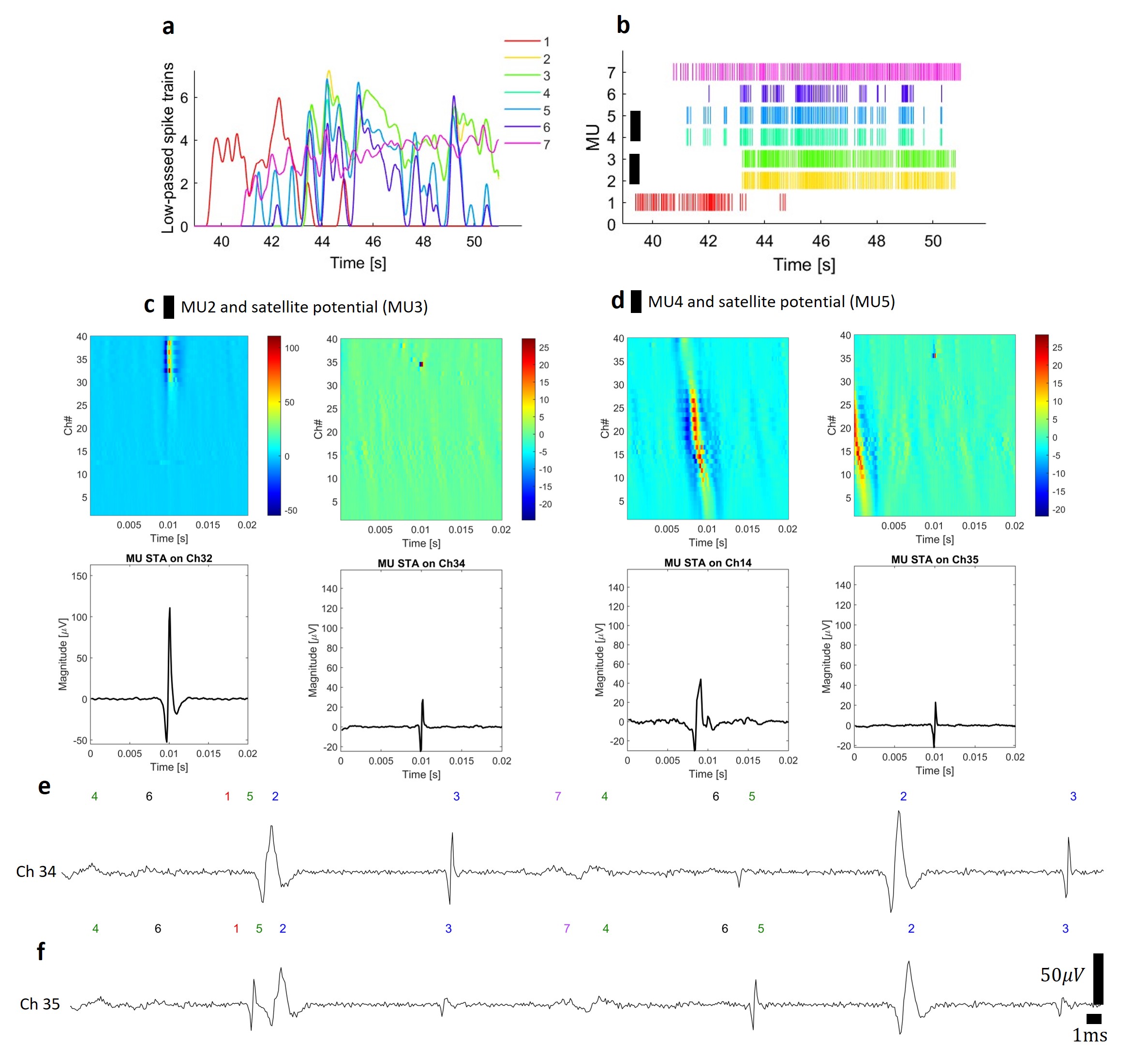}
\caption{Example of satellite potentials. \textbf{a.} Seven motor units are decomposed during trial 4 of Index Extension (TMR4). \textbf{b}. The smoothed spike trains obtained by low-pass filtering the instantaneous discharge rates of individual motor units are shown. Visual inspection of the signals reveals that the spike trains of motor units 2 and 3, and of MUs 4 and 5 matched. The spike trigger average of the EMG signals on each channel provides the 2D image, shown in \textbf{c} (on the left) and \textbf{d} (on the left), of the MU potential distribution across channels of the micro-electrode array. This reveals the presence of the satellite potentials (on the right of panels c and d). \textbf{e.} and \textbf{f} show a portion of EMG signals recorded by channels 34 and 35 of the micro-electrode. Annotation of detected motor units is indicated in correspondence to the instance of time the motor unit fired. The main potential and corresponding satellite potentials can be observed for MU2 and MU4.}\label{fig:thinfilms}
\end{figure}
\newpage
\section{Motor unit tracking and signal stability}\label{sec:supp:robus}
\begin{figure}[h]
\centering
\includegraphics[width=1\textwidth]{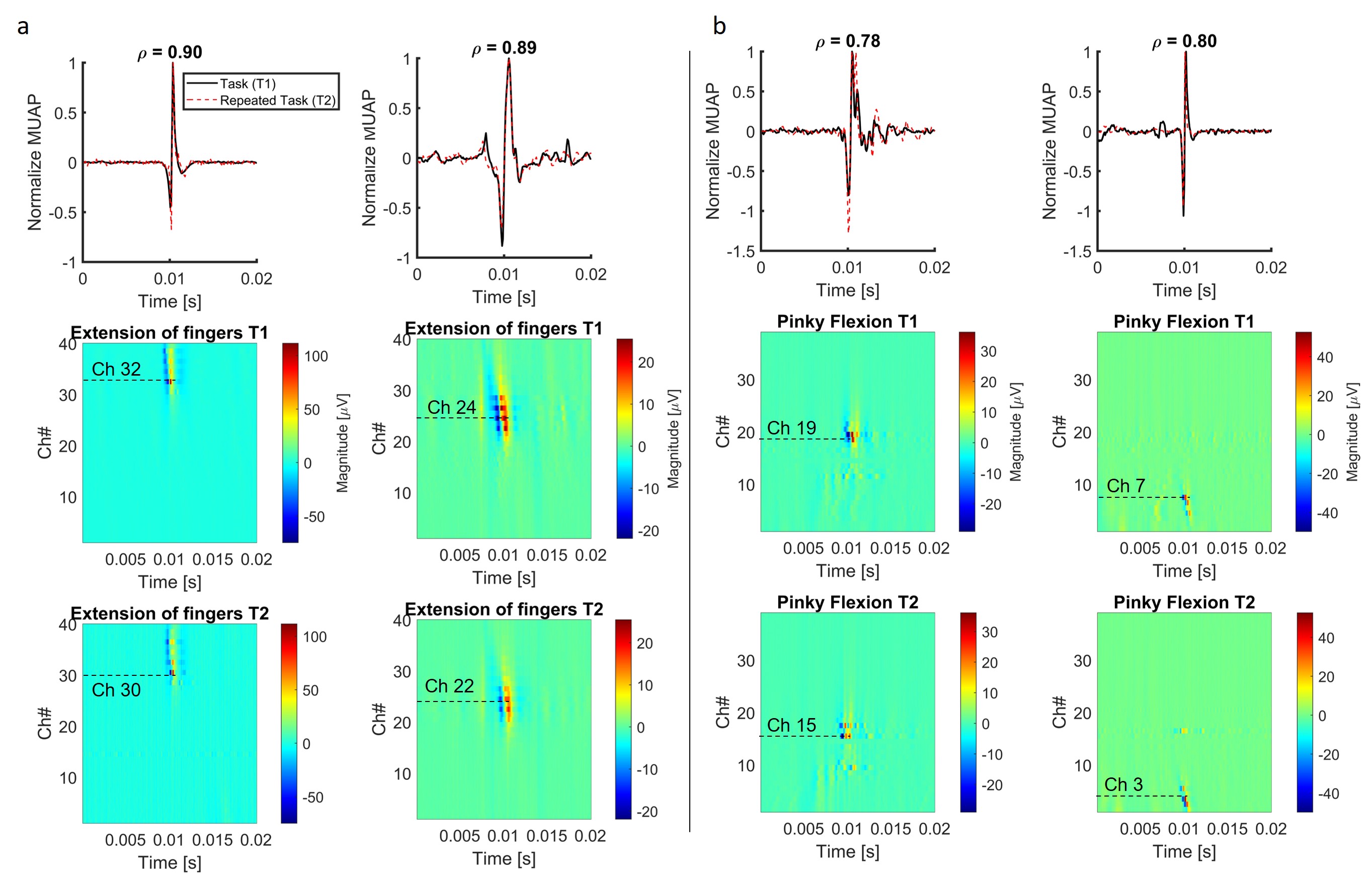}
\caption{Examples of motor units detected during  two instances (beginning T1 and end of experimental session T2) of Extensions of Fingers (i.e., hand opening) \textbf{(a)} and Pinky Flexion \textbf{(b)} for P3.}\label{fig:robust:supp}
\end{figure}

\end{document}